\documentclass[12pt,preprint]{aastex}
\newcommand{\kms}{km\,s$^{-1}$}
\shorttitle{HNCO in Galaxies}
\shortauthors{Mart\'{\i}n et al.}

\begin{document}

\title{HNCO abundances in galaxies: Tracing the evolutionary state of starbursts}
%\title{Tracing heating mechanisms with HNCO: the Galactic center\footnote{Based on observations carried out with the IRAM 30-meter telescope. IRAM is supported by
%INSU/CNRS (France), MPG (Germany) and IGN (Spain).}}

\author{Sergio Mart\'{\i}n}
\email{smartin@cfa.harvard.edu}
\affil{Harvard-Smithsonian Center for Astrophysics, 60 Garden St.,  02138, Cambridge, MA, USA}

\author{J. Mart\'{\i}n-Pintado}
\affil{Departamento de Astrof\'{\i}sica Molecular e Infrarroja, Instituto de Estructura de la Materia, CSIC, Serrano 121, E-28006 Madrid, Spain}
\author{R. Mauersberger}
\affil{Instituto de Radioastronom\'{\i}a Milim\'etrica, Avenida Divina Pastora 7, Local 20, E-18012 Granada, Spain}
\affil{Joint Alma Observatory, Av. El Golf 40, Piso 18, Las Condes, Santiago, Chile}
%\author{M.A. Requena-Torres}
%\affil{Departamento de Astrofis\'{\i}ca Molecular e Infrarroja, Instituto de Estructura de la Materia, CSIC, Serrano 121, E-28006 Madrid, Spain}

\begin{abstract}
The chemistry in the central regions of galaxies is expected to be strongly influenced by their nuclear activity.
To find the best tracers of nuclear activity is of key importance to understand the processes taking place in the
most obscured regions of galactic nuclei.
In this work we present multi-line observations of CS, C$^{34}$S, HNCO and C$^{18}$O in a sample of 11 bright
galaxies prototypical for different types of activity.
The $^{32}$S/$^{34}$S isotopic ratio is $\sim10$, supporting the idea of
an isotopical $^{34}$S enrichment due to massive star formation in the nuclear regions of galaxies.
Although C$^{32}$S and C$^{34}$S do not seem to be significantly affected by the activity type,
the HNCO abundance appears highly contrasted among starburst.
We observed HNCO abundance variations of nearly two orders of magnitude.
The HNCO molecule is shown to be a good tracer of the amount of molecular material fueling the starburst
and therefore can be used as a diagnostics of the evolutionary state of a nuclear starburst.
%which can be used to define a sequence in the nuclear starburst evolution.
\end{abstract}


\section{Introduction}
The nuclear regions of active galaxies can harbor different energetic phenomena such as starbursts (SBs) and Active Galactic Nuclei (AGN) which
are responsible for producing the bright emission stemming from their central regions.
The heating of the interstellar medium (ISM) in nuclei with different type of activity is expected to be dominated by a variety of
mechanisms, such as UV, X-rays, and/or shocks, which will ultimately drive their observed chemical richness \citep[][and references therein]{Martin06b}.
Disentangling which of this mechanisms is the main power source of galactic nuclei is complicated by the high obscuration
affecting these objects. This fact becomes critical in the case of Ultra Luminous Galaxies (ULIRGs) and high-z sources.
Therefore the observation of the dense molecular material
%, intimately related to star formation locations,
has become an essential tool to get an insight in the evolution and classification of the nuclear activity in galaxies.

Finding appropriate molecular species to accurately trace each of these heating mechanisms is
%the key to achieve this aim.
crucial to stablish the nature of the central engine.
Molecular species like HCN and HCO$^+$, and in particular the ratios HCN/CO and HCN/HCO$^+$, have been claimed to be appropriate to differentiate
between the SB and the AGN contribution in galactic nuclei \citep{Kohno99,Kohno05,Krips08}.
Similarly, species such as HNC and CN are thought to show
enhanced abundances in extremely irradiated environments 
%\citep[PDRs and XDRs][]{Aalto02,Aalto07}.
\citep{Aalto02,Aalto07}.
A dichotomy still remains concerning the different evolutionary state of nuclear starburst.
While at an early stage, the heating of the ISM is thought to be dominated by shocks affecting the molecular clouds fueling the starburst, the late
stages of starburst are vastly dominated by the UV radiation originated in the newly formed massive stars.
This scenario has been inferred from the extensive observation of the prototype sources NGC\,253 and M\,82 in a number of key molecular tracers
such as SO$_2$, H$_2$S, OCS \citep{Martin03,Martin05}, HOC$^+$, and CO$^+$ \citep{Fuente05,Fuente06}.
Unfortunately, these tracers are generally too faint to be detected, but for the brightest prototypical nearby galaxies.

In a recent observational study carried out towards a sample of molecular clouds dominated by different heating mechanisms 
%chemistry prototypes 
within the Galactic center region \citep{Martin08},
we have found that the CS/HNCO abundance ratio is highly contrasted between molecular clouds illuminated by the UV radiation from massive star
clusters and the giant molecular cloud complexes shielded from photodissociation and mostly heated by shocks.
The origin of the large changes in the CS/HNCO ratio is due to the enhancement of CS in photon-dominated regions (PDRs) through reactions
involving S$^+$ \citep{Stern95} as opposed to the highly
photodissociable HNCO, which is destroyed by the UV photons.
Both observations of photodissociation regions within our Galaxy and photochemical models show the suitability of CS as a PDR tracer \citep{Goico06}.
In addition, HNCO is enhanced in the presence of shocks due to its injection in the gas phase from the grain mantles \citep{Zinchen00}.
This paper presents a follow up study of the CS/HNCO ratio towards a sample of prototype galaxies with different nuclear activity.
Although CS has been extensively observed in external galaxies \citep{Baan08},
HNCO had only been detected in four galaxies prior to this work \citep{Nguyen91,Wang04}.

\section{Observations and Results}

Observations were carried out with the IRAM 30\,m telescope on Pico Veleta, Spain,
during three observing periods from summer 2005 through summer 2007.
Table~\ref{tab:sourceandlines} shows the sample of galaxies with their coordinates and the
rest frequencies of the observed molecular transitions.
Observations were performed in symmetrical wobbler switched mode with a frequency of 0.5\,Hz and a beam throw of
$4'$ in azimuth.
As spectrometers we used the $512\times1$\,MHz filter banks for the 3\,mm transitions and the $256\times4$\,MHz filter banks for
those at 2 and 1.3\,mm.
Pointing accuracy was estimated to be of $\sim3''$ from frequent continuum cross scans on nearby pointing sources.
Data were calibrated using the standard dual load system and main beam temperatures were obtained as
$T_{\rm MB}=(F_{\rm eff}/B_{\rm eff})T^*_{\rm A}$, where $B_{\rm eff}$ is tabulated in Table~\ref{tab:sourceandlines} and $F_{\rm eff}$
are 0.95, 0.93, and 0.91 for 3, 2, and 1.2\,mm, respectively. Calibration uncertainties are estimated to be $<15\%$.

\begin{figure}
%\epsscale{1.3}
\includegraphics[angle=0,scale=0.29]{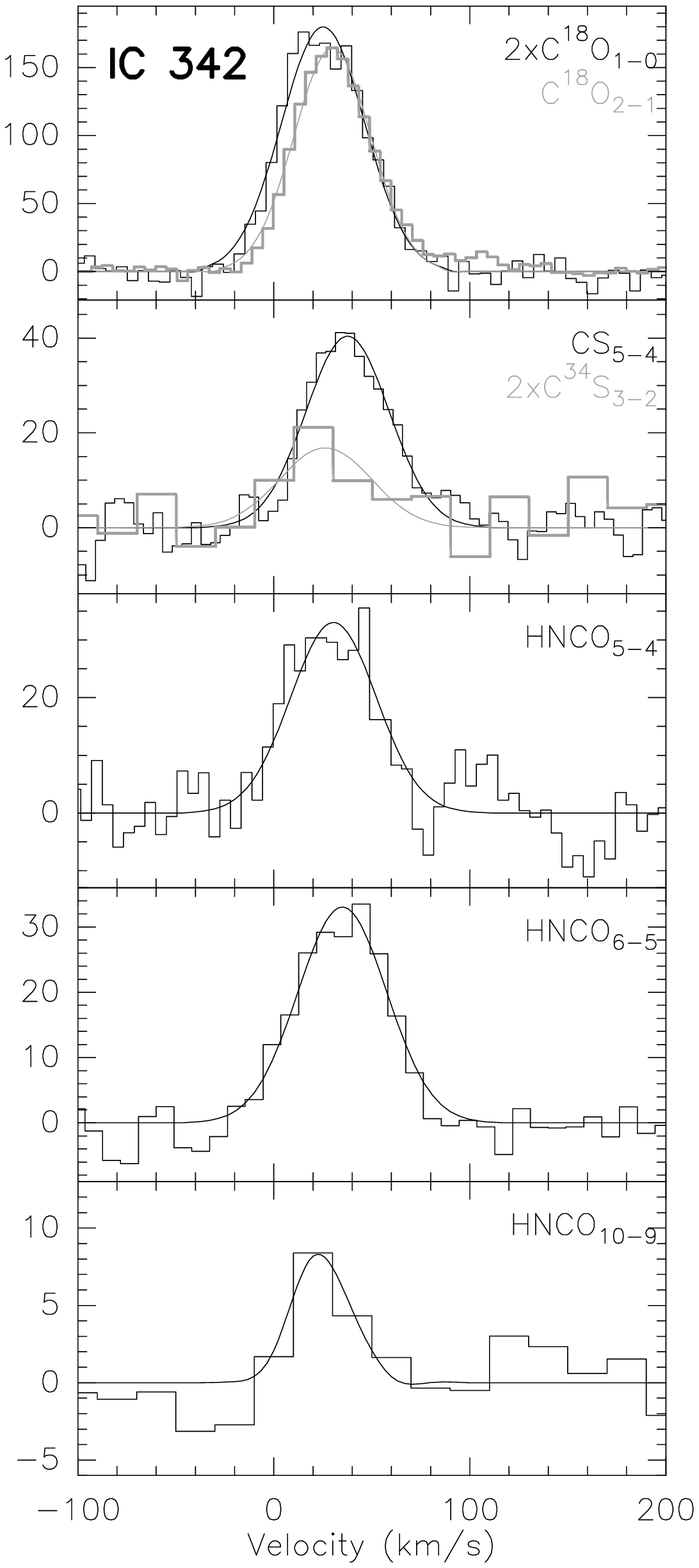}
\includegraphics[angle=0,scale=0.29]{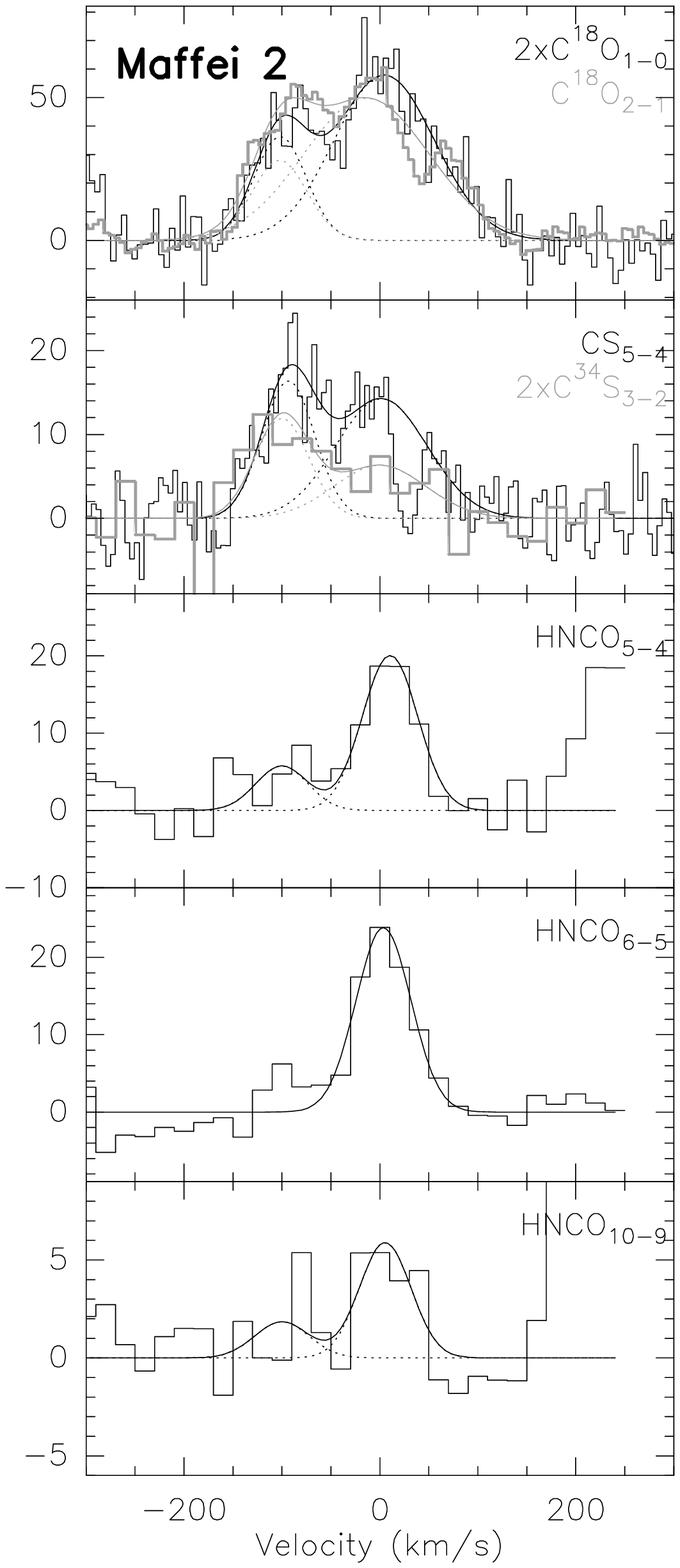}
\includegraphics[angle=0,scale=0.29]{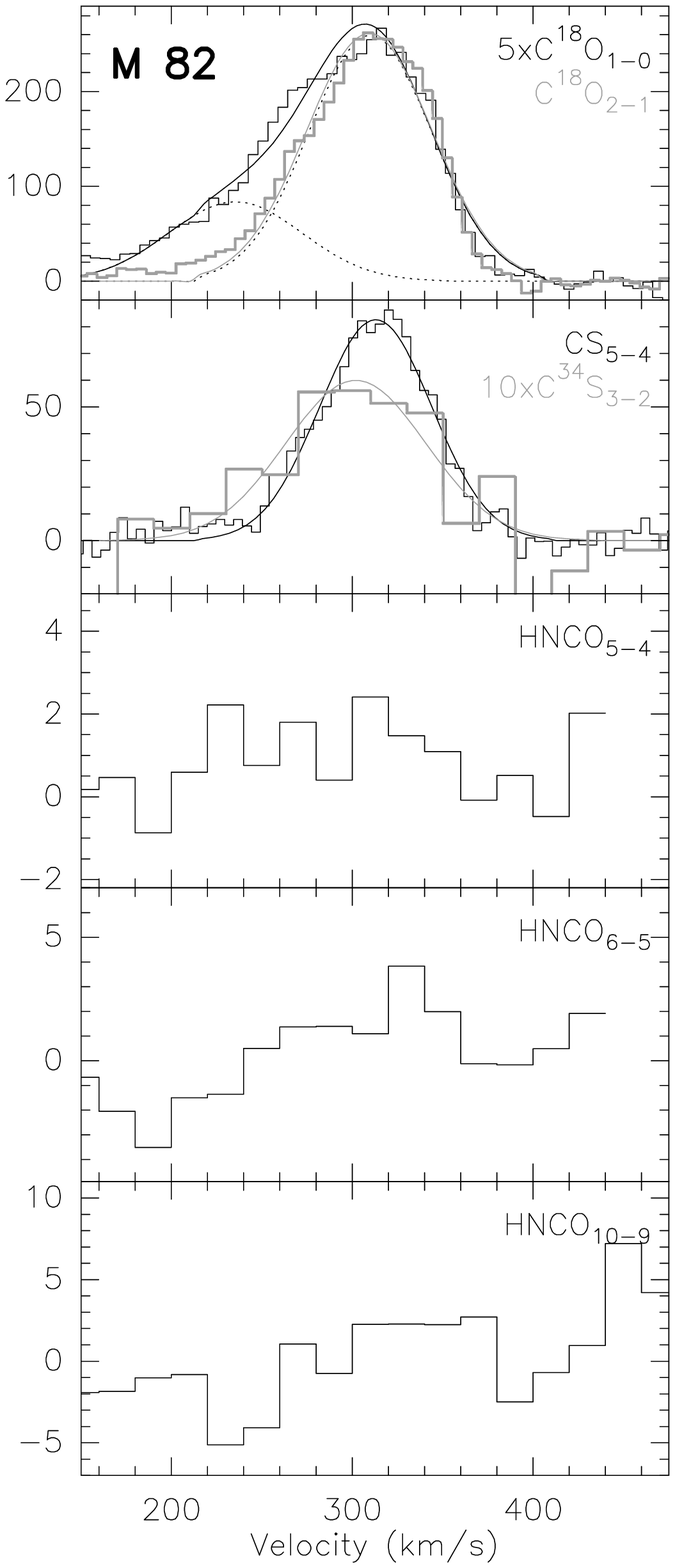}
\includegraphics[angle=0,scale=0.29]{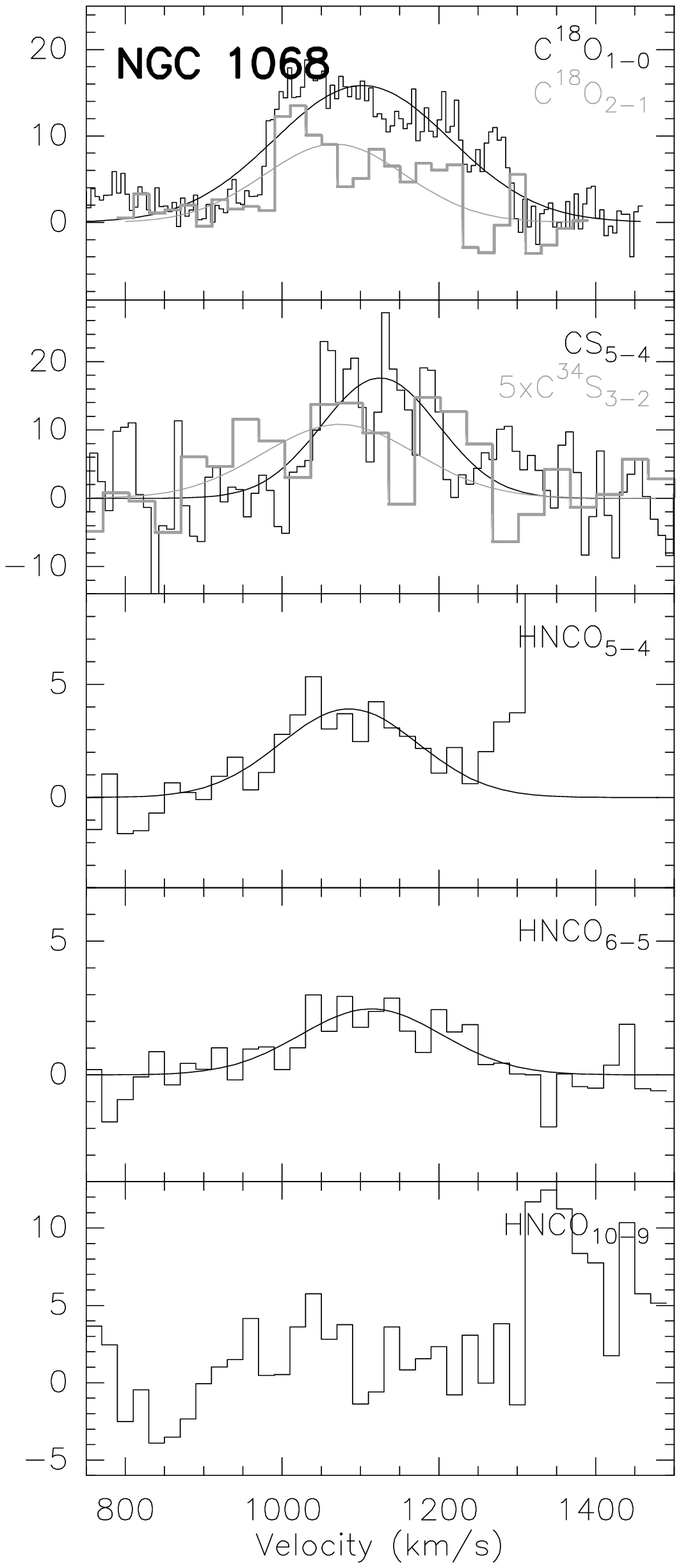}
\includegraphics[angle=0,scale=0.29]{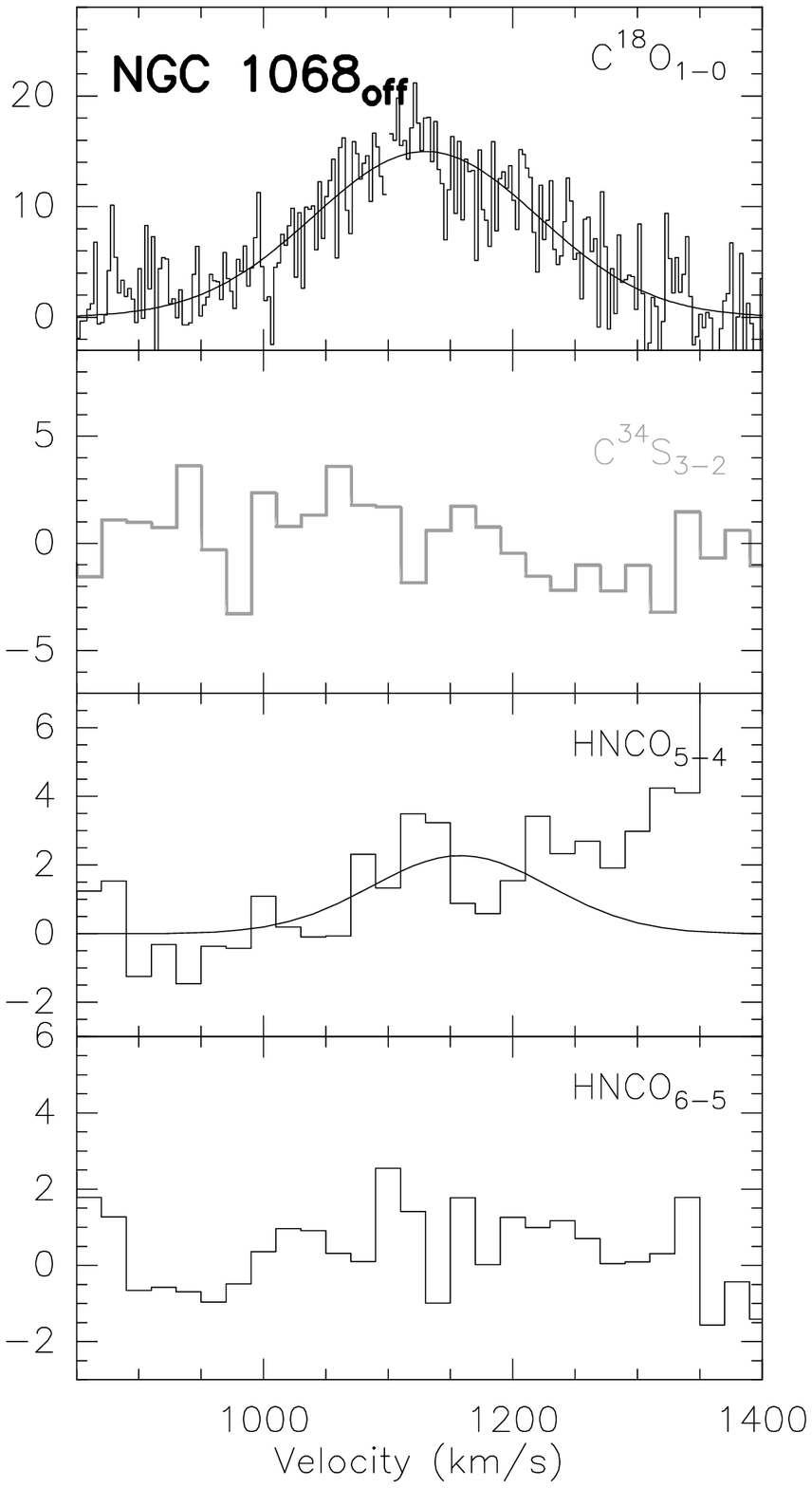}
\includegraphics[angle=0,scale=0.3]{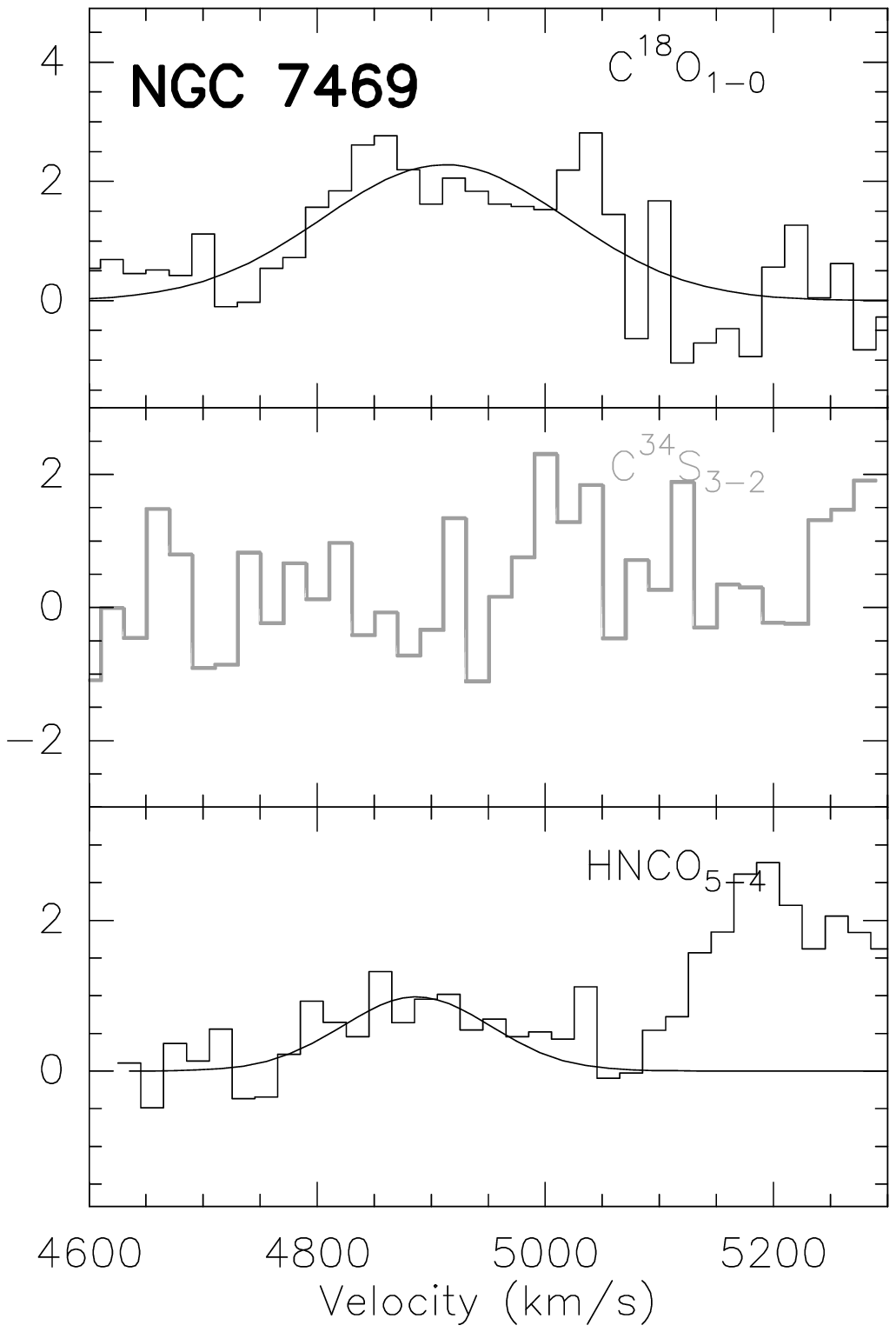}
\includegraphics[angle=0,scale=0.3]{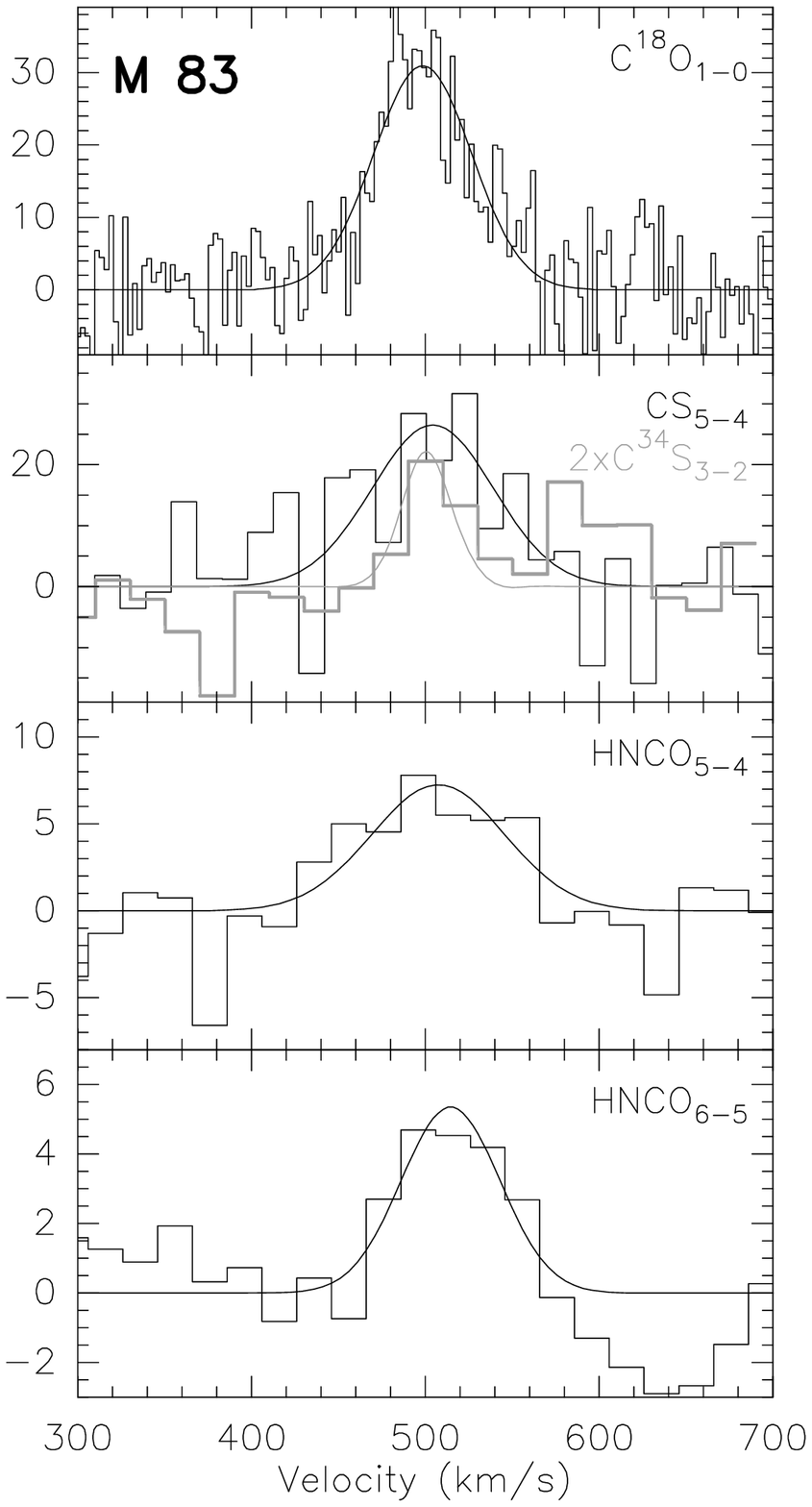}
\includegraphics[angle=0,scale=0.3]{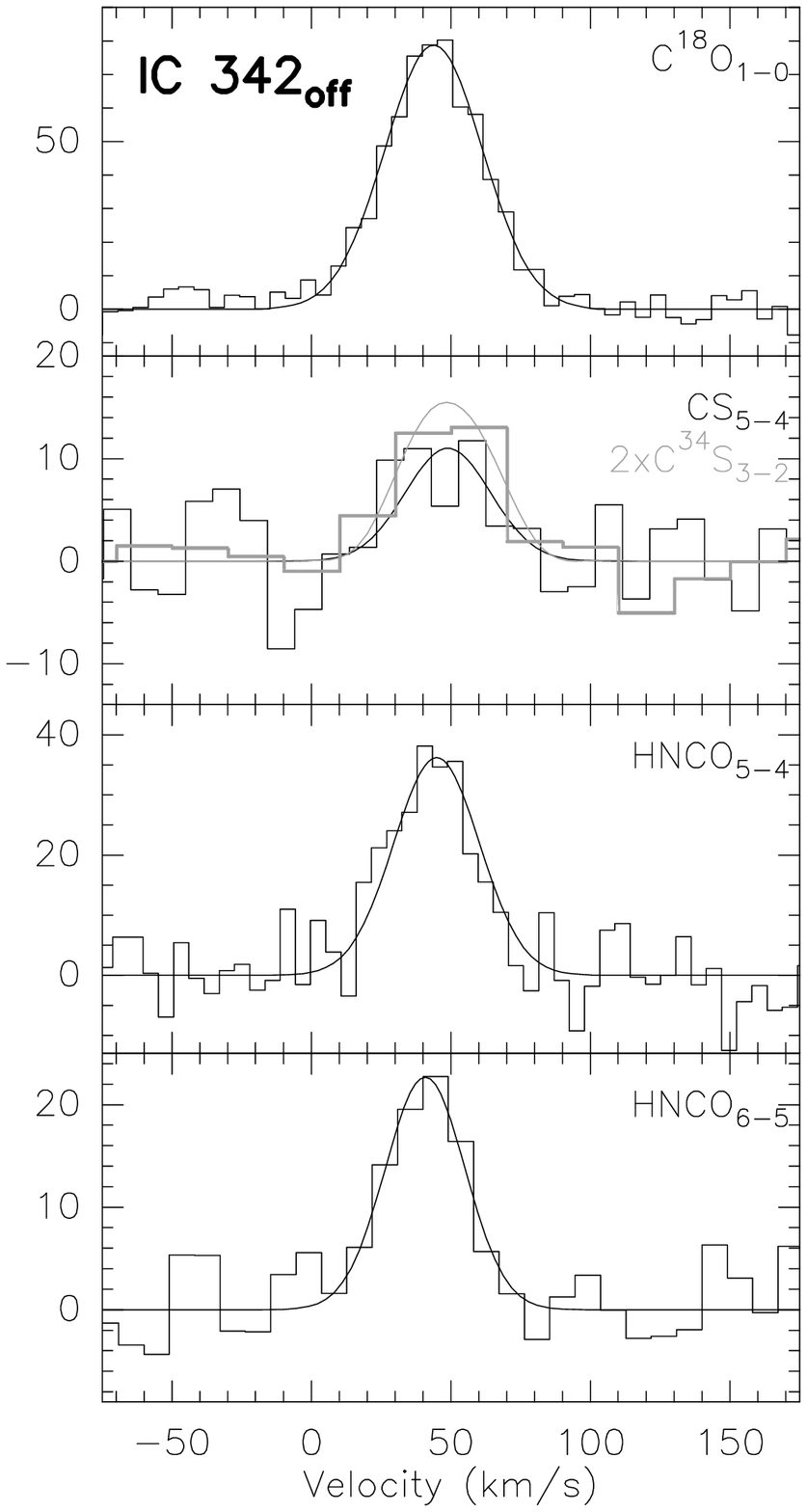}
\includegraphics[angle=0,scale=0.3]{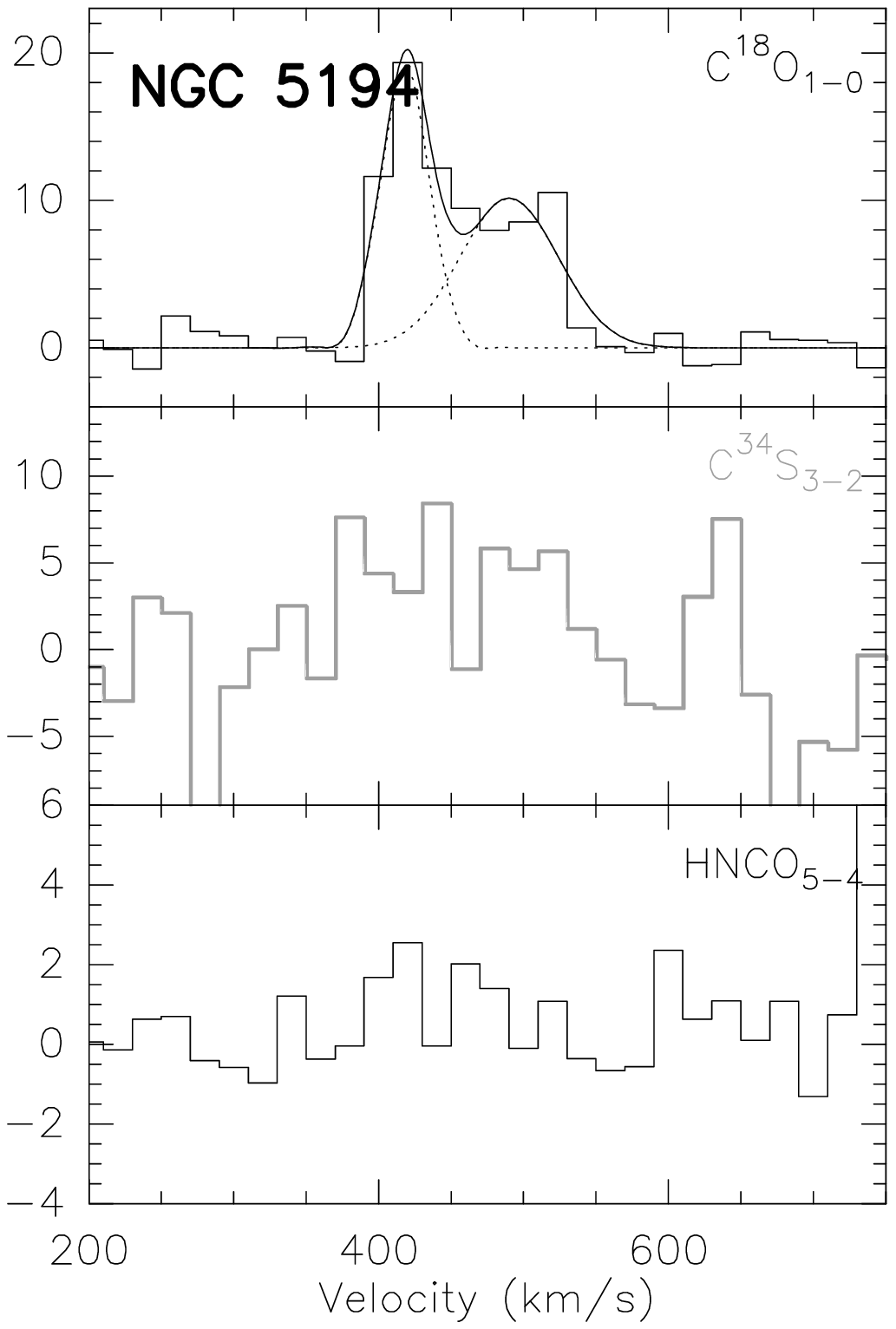}
\includegraphics[angle=0,scale=0.3]{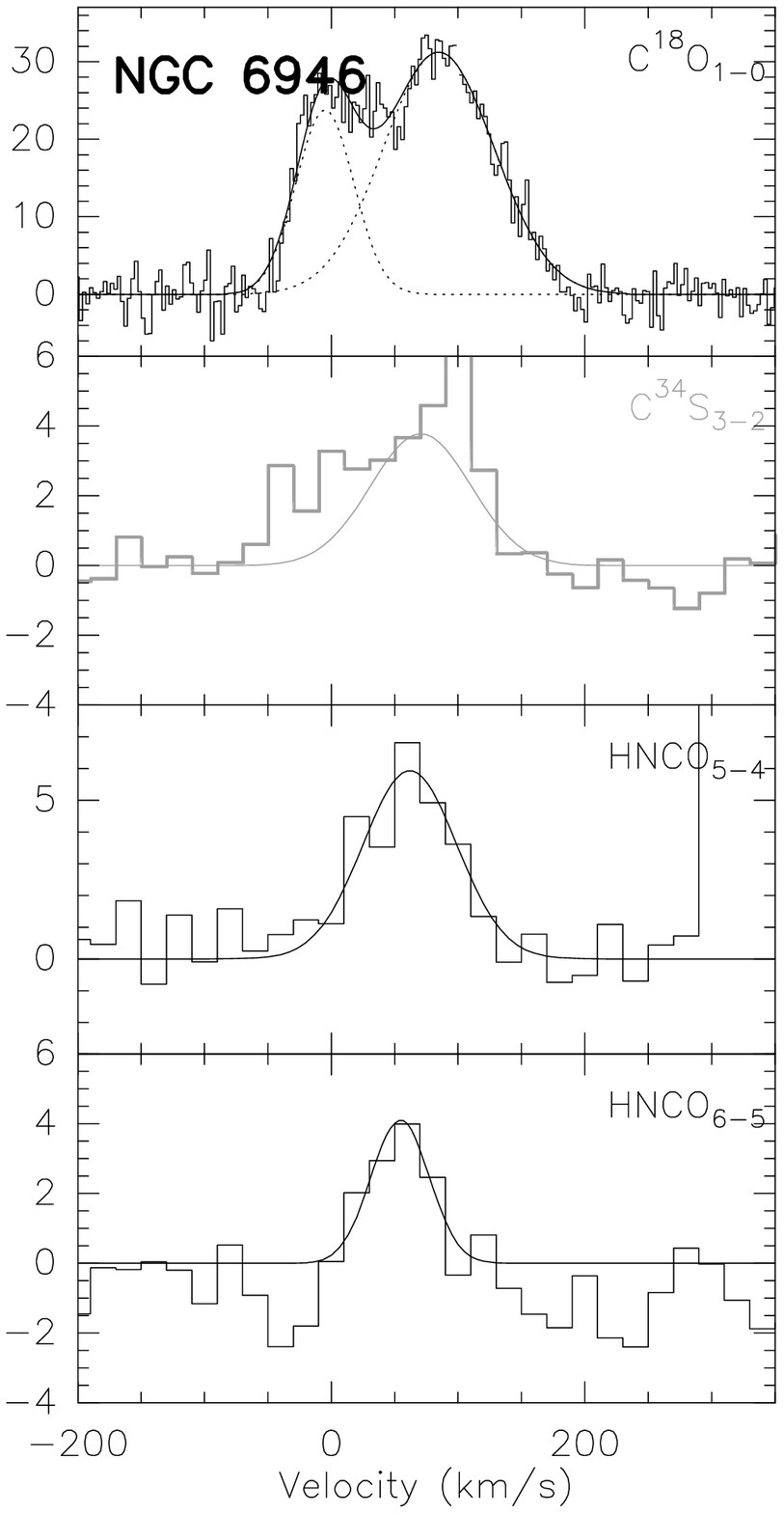}
\includegraphics[angle=0,scale=0.3]{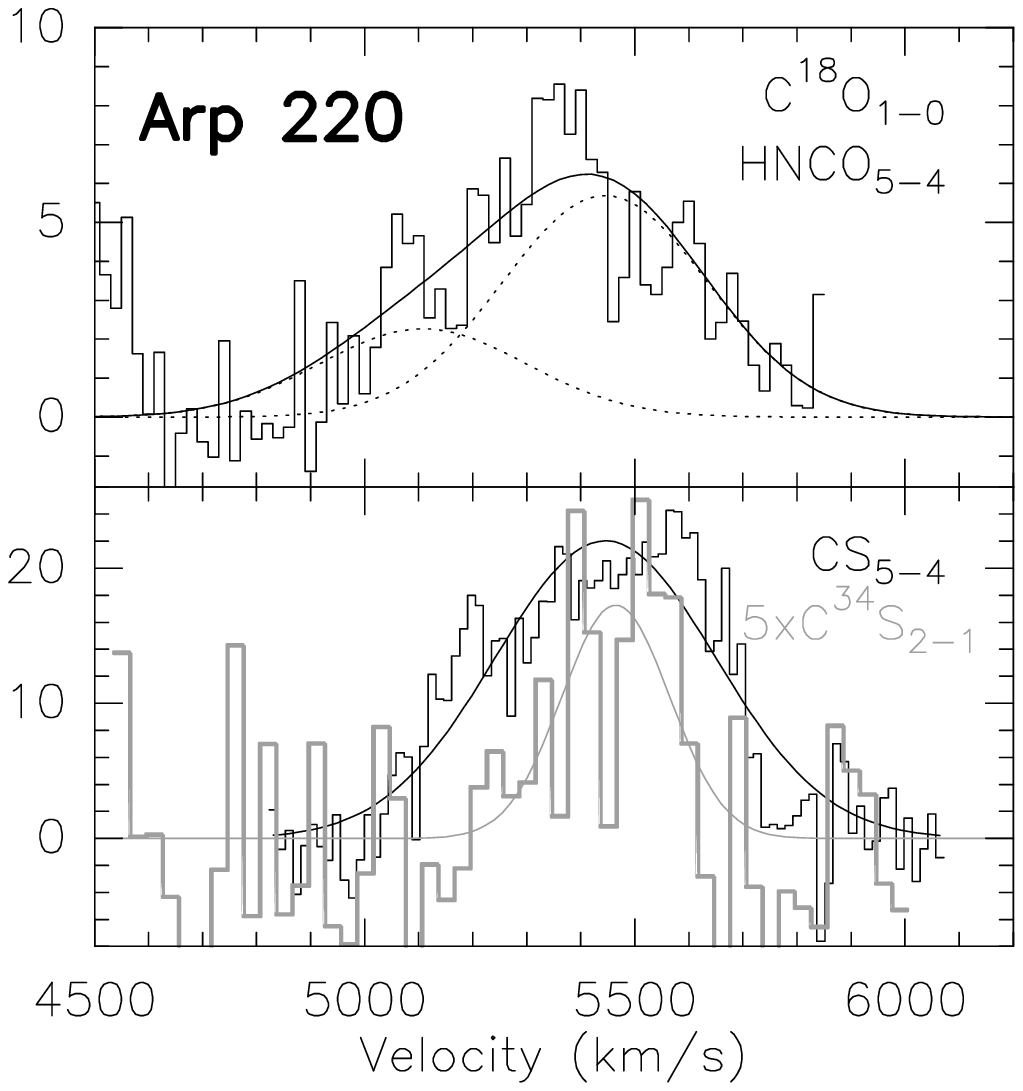}
\caption{Observed spectra for each source in the sample.
Some of the transitions C$^{18}$O $1-0$ and C$^{34}$S $3-2$ spectra have been scaled up for the sake of clarity in the profile comparison.
Spectra with line intensities below 10\,mK have been resampled to a velocity resolution of 20\,\kms, otherwise original velocity
resolution is shown.
C$^{34}$S\,$(3-2)$ in NGC\,1068 and C$^{34}$S\,$(2-1)$ in Arp\,220 are shown with a 30\,\kms velocity resolution.
Gaussian fits are shown for all detected transitions. Dotted fits indicate individual velocity components.
For Arp\,220, where C$^{18}$O and HNCO transitions appear significantly blended, 
we show both fits in the same spectrum where the overall fit is shown with a continuous line and the individual transition
with dotted lines.
Temperature scale is $T_{\rm MB}\,(\rm mK)$.}
\label{fig.spectra}
\end{figure}

The observed spectra for each source are shown in Fig.~\ref{fig.spectra}.
The HNCO $5-4$ and $10-9$ lines were observed simultaneously with the C$^{18}$O $1-0$ and $2-1$ lines, respectively.
This allows a reliable determination of the HNCO relative abundance, as no uncertainties due to pointing drifts and/or calibration
are involved.
For some of the sources with the larger observed linewidths, the HNCO and C$^{18}$O lines appear moderately blended.
Such is the case of NGC\,7469, NGC\,1068, and Arp\,220.
Gaussian shapes have been fitted to the observed profiles. 
Table~\ref{tab.gaussfit} compiles the derived fitted parameters.
For the non-detected transitions, upper limits are tabulated where it was considered a $3\sigma$ level and an approximate linewidth
similar to those of the detected CS and HNCO transitions, or that of C$^{18}$O in the case of NGC\,5194.

Rotational temperatures and molecular column densities have been derived assuming local thermodynamic equilibrium (LTE) and
optically thin emission.
As derived for CS $J=3-2$ in NGC\,253 \citep{Martin05} we only might expect the emission of CS $J=5-4$ to be significantly optically thick.
However, our observation of C34S will be less affected by the opacity effects.
We have accounted for the dilution of the source in the telescope beam by assuming an averaged emission extent of $10''$ for all sources.
This simplification might affect the determination of the column densities for sources much smaller than $10''$ 
\citep[see][]{Martin06b}, but the abundance ratios between species dealt with in this paper (Sect.~\ref{sec.CSvsHNCO})
will not be significantly affected.
Thus, the main source of uncertainty will be due to possible different emission extents between molecules.
Table~\ref{tab.coldens} shows the column density and rotational temperature derived for every source and species.
From the LTE analysis, we derive C$^{18}$O rotational temperatures ($T_{\rm rot}$) of $3-7$\,K for most of the galaxies except for M\,82
where we determine $T_{\rm rot}$=15\,K.
The temperatures derived from the HNCO line emission of $4-18$\,K are a factor of $\sim 2$ higher than those derived from C$^{18}$O.
Whenever only one transition was observed, the average rotational temperature of 10\,K was assumed.

We have also included the available observations towards NGC\,253 to derive its physical parameters shown in Table~\ref{tab.coldens}
\citep{Harrison99,Martin05,Martin06b}.
Note that the C$^{18}$O transitions in NGC\,253 were observed $\sim10''$ north of the position shown in Table~\ref{tab:sourceandlines}.
For NGC\,4945 we used the observations by \citet{Wang04}.
Not all sources in the sample were observed in the CS $J=5-4$ transition.
The upper limit to the emission to this transition in NGC\,6946 \citep{Mauers89} allows to constrain the CS column density to $<6.2\times10^{13}\rm cm^{-2}$.
We have thus used the CS $J=3-2$ transitions available on NGC\,6946 and NGC\,5194 \citep{Mauers89} to estimate the column densities presented in Table~\ref{tab.coldens}.
Note that for NGC\,6946 we used the observed offset ($\Delta\alpha=10'',\Delta\delta=-10''$) position in \citet{Mauers89}, which is located just $\sim1''$ away from our nominal position.
Similarly, their NGC\,5194 ($0'',0''$) position is $\sim4''$ away from our observed position.

The critical densities for CS and HNCO transitions observed in this works are expected to be very similar since the Einstein coefficients and collisional
cross sections are estimated to be similar \citep{Schoier05}.
Thus, the HNCO/CS ratio should reflect a reliable abundances ratio.
In any case, the possible uncertainties introduced by slightly different critical densities will never account for the large difference in the abundance ratios
measured in different galaxies.
The differences in the derived molecular abundances might also be affected by molecular excitation differences of each individual source.
%s not a real difference in the chemical compositionthe observed sources.
However the detailed multitransition study and LVG modeling of dense gas tracers such as HCN and HCO+ in a sample of galaxies
by \citet{Krips08}, shows how for several molecular transitions with similar critical densities, 
the derived abundances from the line intensities are really determined by real differences in abundances and not by excitation effects.
The low excitation temperatures derived from C$^{18}$O indicates that a substantial fraction of the column density will be dominated by relatively low density gas.
Since in our diagnostic diagrams in Fig.~\ref{fig.CSHNCO} are normalized to the total column densities derived from C$^{18}$O,
changes in the H$^2$ total column densities will not change the main conclusions of the paper.

\section{Selected sources}
The sample of sources observed in this paper has been selected among the brightest prototypes for the different types of nuclear activity.
Their distances and corresponding linear scales are shown Table~\ref{tab:sourceandlines}.
The sources can be grouped as follows.
%starburst galaxies (SBGs) and Active Galactic Nuclei (AGNs).

\subsection{Starburst Galaxies (SBG)}
{\it M\,82} and {\it NGC\,253} are the strongest and richest extragalactic molecular sources.
These are the archetypes of starburst galaxies housing the two brightest extragalactic IRAS sources \citep{Soifer89}.
M\,82, as opposed to the young starburst in NGC\,253, is claimed to be the prototype of evolved starburst.
This is supported by its observed chemistry characterized by low abundance of complex molecules such as SiO, CH$_3$OH \citep{Mauers93,Martin06b} 
%while strong irradiation tracers as 
and large abundance of molecular ions like CO$^+$, HOC$^+$, and H$_3$O$^+$ which are expected to be enhanced in PDRs \citep{Fuente05,Fuente06,vdtak08}.
In this work we did not observe the central position towards M\,82, but a position in the north-east molecular complex (hereafter M\,82$^*$)
where most of the photodissociation regions are located as observed in the HCO emission high-resolution maps \citep{Burillo02}.
{\it Maffei\,2}, is another well studied starburst spiral galaxy with a high molecular gas concentration in its nuclear region
and showing traces of a tidal interaction with a dwarf companion galaxy \citep{Hurt96,Mason04}.
{\it M\,83} is a nearby face on barred galaxy undergoing strong nuclear starburst likely due to gas inflow along its bar \citep{Petipas98,Talbot79}.
{\it NGC\,6946} shows a moderate starburst in its nucleus also claimed to be related to the presence of a bar \citep{Turner83,Schinne07}.
% Previous bursts 7 Myr and 20Myr Engelbracht, C. W., Rieke, M. J., Rieke, G. H., & Latter, W. B. 1996

\subsection{Active Galactic Nuclei (AGN)}
{\it NGC\,1068} is a nearby luminous infrared galaxy (LIRG) prototype of a Seyfert 2 nucleus \citep{Anto85}.
The AGN is enclosed by a circumnuclear starburst ring with $14''$ ($\sim1$\,kpc) radius \citep{Myers87,Schinne00}.
Thus, we have observed two positions in NGC\,1068. One towards the central AGN and an offset position $0'',-14''$ towards a peak of emission within the 
circumnuclear ring, hereafter NGC\,1068$^*$, mostly tracing the starburst.
{\it NGC\,4945}, being one of the three brightest IRAS point sources, has been the target of some a detailed molecular study by \citet{Wang04}.
This nearly edge-on spiral galaxy harbor a heavily obscured Seyfert 2 nucleus \citep{Braatz97,Maio99} also surrounded by a starburst ring \citep{Marconi00}.
{\it NGC\,5194} (M\,51) is a nearby grand-design spiral galaxy with a Seyfert 2 nucleus.
The proximity of this galaxy as well as its nearly face-on orientation has made it the target of numerous large scale multi-wavelength
studies \citep{Scoville83,Scoville01,Schuster07}.
Its interaction with the companion galaxy NGC\,5195 seems to have been the origin of a past period of intense star formation \citep{Thronson91,Greenawalt98}.
{\it NGC\,7469}, the only example of Seyfert 1 nucleus in our sample, also shows a circumnuclear star-forming ring of $3''$ in diameter \citep{Wilson91}.

\subsection{UltraLuminous Infrared Galaxies (ULIRG)}
{\it Arp\,220} is one of the best studied nearby ULIRGs \citep{Dopita05} with more than a 95$\%$ of its luminosity radiated at
IR/submm wavelengths \citep{Sanders03}.
At a distance of  D$\sim77$\,Mpc ($z=0.018$) and star formation rates of 300$\,M_\odot$\,yr$^{-1}$, Arp\,220 is often referred as 
a nearby template of the luminous star-forming galaxies at high-z.
Arp\,220 is the result of an advanced merger system with two nuclei separated by $\sim1''$ ($\sim300$\,pc) as 
seen from near-IR and radio images \citep{Baan95,Scov00}
with an enormous concentration of molecular gas within its central region  \citep[a few $10^{9}\,M_\odot$,][]{Saka99}.

\subsection{``Normal''}
We refer to {\it IC\,342} is as a ``normal'' galaxy as being the closest spiral galaxies resembling our own Galaxy. 
It shows a minispiral structure meeting at the inner molecular ring surrounding the central stellar cluster \citep{Ishizuki90,Boker97,Helfer03}.
Chemical differences have been observed between the gas in the nuclear region, affected by the intense star formation through the radiation from the central
cluster, and the molecular material in the spiral structure, mainly affected by shocks \citep{Meier05,Usero06}.
In this work we have observed two positions, one at the central region and the other at
the offset position $(+5'',+15'')$, hereafter {\it IC\,342\,$^*$}, located just in one of the arms of the minispiral.

\section{Discussion}

\subsection{$\bf C^{32}S$ and $\bf C^{34}S$}
We have derived the abundances of CS, C$^{34}$S, and HNCO relative to H$_2$.
In order to estimate the H$_2$ column density we have used the observed C$^{18}$O emission, assuming a $^{16}$O/$^{18}$O abundance ratio of 150, as derived 
in NGC\,253 \citep{Harrison99} and the standard CO/H$_2$ ratio of $10^{-4}$.

The CS fractional abundance shows a narrow range of values within $1-3\times10^{-9}$.
Similarly the C$^{34}$S abundances are concentrated in the range of $1-5\times10^{-10}$.
There are only three exceptions to this trend, namely IC\,342$^*$, Arp\,220, and NGC\,5194.
The abundances of CS observed towards IC\,342$^*$ is a factor of 6 below the average in the sample, while the C$^{34}$S abundance is close to the
average value. This is explained by the highly concentrated CS emission towards the nuclear region \citep{Meier05} and the almost twice smaller
beam size at the CS\,$5-4$ transition frequency compared to that at the C$^{34}$S\,$3-2$ line (see Table~\ref{tab:sourceandlines}).
We did not detect neither CS nor C$^{34}$S emission towards NGC\,5194. While the C$^{34}$S upper limit is still within the average value obtained for the
other galaxies, the limit we derive in the main isotopologue is a factor of 3 below the sample average.
%The galaxy NGC\,5194 shows a significantly lower abundance of CS. The upper limit to the C$^{34}$S abundance is not low enough to confirm whether
%it shows the same low relative abundance as the main isotopologue.
On the other hand, the galaxy Arp\,220 is observed to have an abundance a factor of $\sim 5$ over the average in both isotopologues.
This points out a real overabundance of CS towards this ULIRG.

\begin{figure}
\includegraphics[angle=0,scale=0.8]{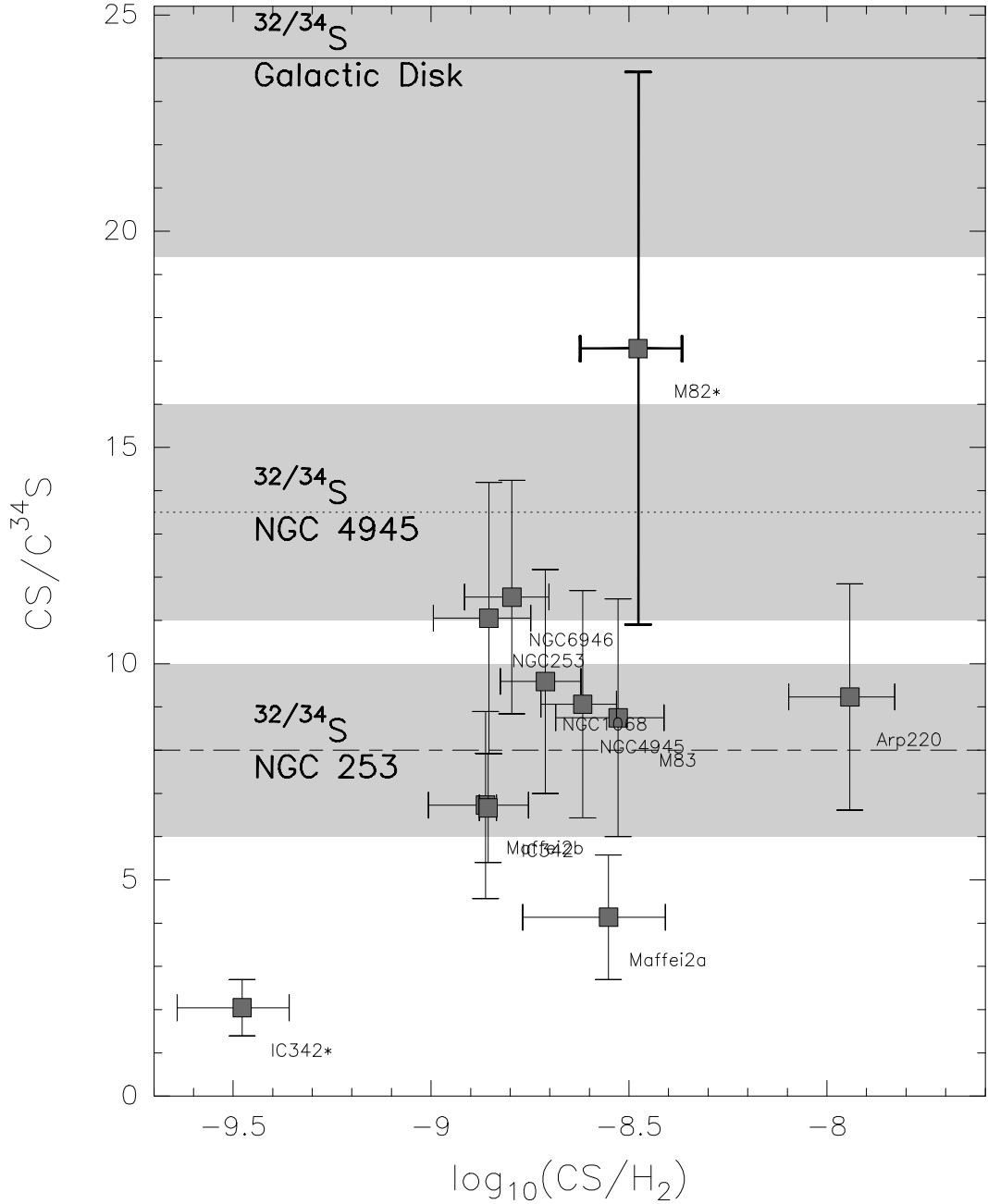}
\caption{C$^{32}$S/C$^{34}$S Abundance ratio for the detected galaxies as a function of the CS fractional abundance.
Sulfur isotopic ratios from the literature for the Galactic Disk, and the galaxies NGC\,4945 and NGC\,253 are represented with
continuous, dotted and dashed lines, respectively, and their corresponding error shown as a grey shade.
Labeling of sources is the same as in Fig.~\ref{fig.CSHNCO}.}
\label{fig.CSC34S}
\end{figure}

Fig.~\ref{fig.CSC34S} shows the C$^{32}$S/C$^{34}$S abundance ratio versus the fractional abundance of CS.
This ratio is calculated with the 1\,mm CS $5-4$ and the 2\,mm C$^{34}$ $3-2$ transitions which makes it dependent on the
assumed rotational temperature to derive the column densities, and to a lesser extent (up to a 30\% for the smallest sources)
on the considered source size.
However, the ratio derived for NGC\,4945 and NGC\,253 agree within a factor of 1.5 with the values derived from the complete study of CS by
\citet{Wang04} and \citet{Martin05}, respectively.
%Indeed, NGC\,4945 and NGC\,253 are the only two galaxies where the isotopical ratio $^{32/34}$S has been derived.
Although the effects of line opacity have not been taken into account, we can consider the derived C$^{32}$S/C$^{34}$S abundance ratio
as a rough representation of the $^{32/34}$S isotopic ratio.
It appears that almost all the sources in the sample have ratios around $\sim10$, in agreement with those previously derived for NGC\,4945 and NGC\,253,
and well below the value of $\sim 24$ measured in the Galactic disk \citep{Chin96}.
This result supports the idea of an $^{34}$S overproduction in the nuclei of galaxies by massive stars.
Furthermore, we observe this ratio to be independent of the type of nuclear activity, both in AGN and SB dominated galaxies.

\subsection{CS versus HNCO abundances}
\label{sec.CSvsHNCO}

\begin{figure}
\includegraphics[angle=0,scale=0.8]{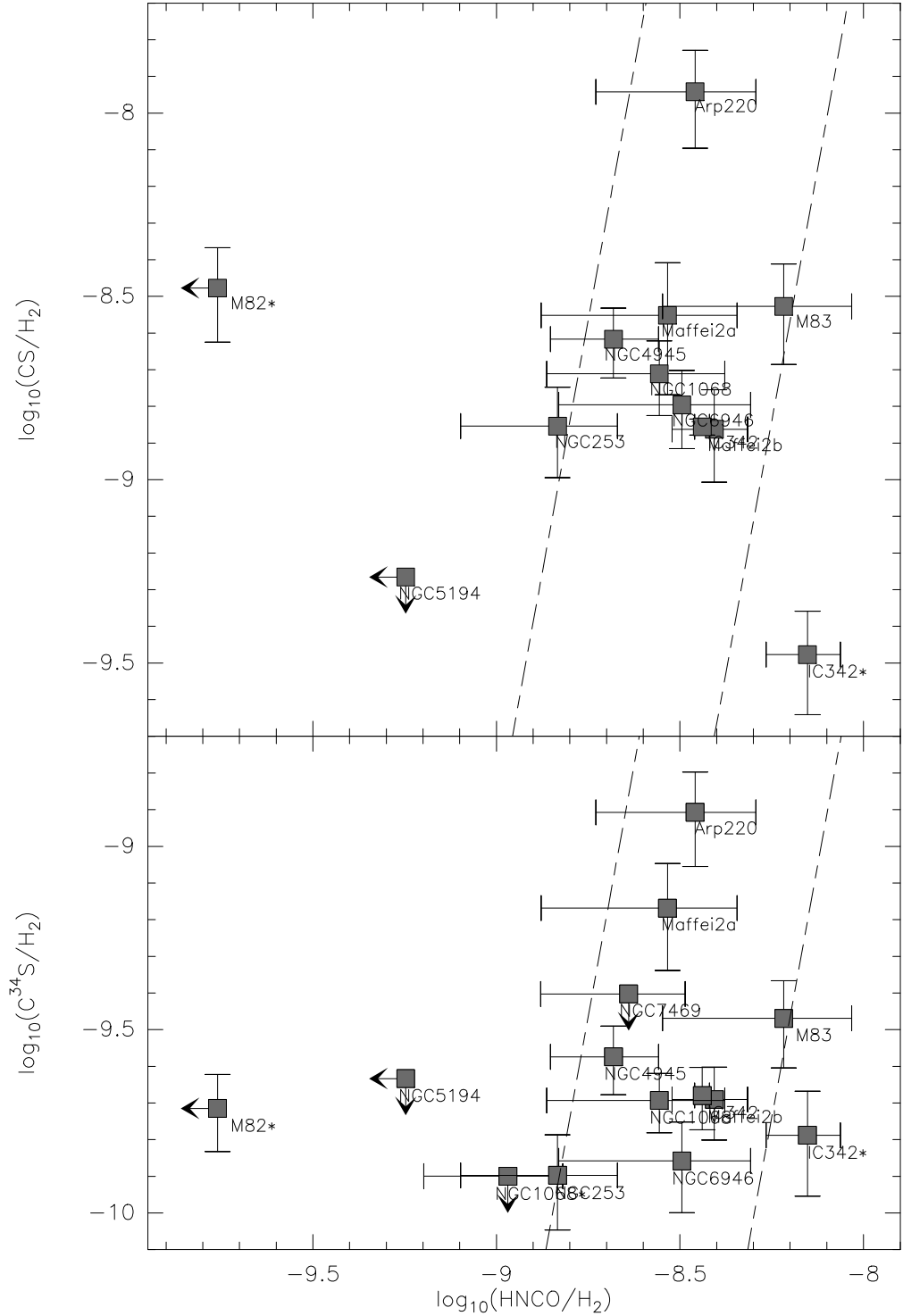}
\caption{Fractional abundances of CS ({\it Top}) and C$^{34}$S ({\it Bottom}) vs the fractional abundance of HNCO with respect to H$_2$.
Upper limits to the abundances are represented by arrows.
As in the text, offset observed position are labeled with an * following the galaxy name.
The two velocity components fitted in Maffei2 are labeled as Maffei2a and Maffei2b.
Dashed lines represent the limits of the regions used to discriminate the UV dominated (left side) from the shock dominated (right side) molecular
gas, as derived from Galactic center observations \citep[see][for details]{Martin08}.}
\label{fig.CSHNCO}
\end{figure}

In Fig.~\ref{fig.CSHNCO}, we represent the fractional abundances of CS and C$^{34}$S versus those of HNCO.
We observe that most of the galaxies in the sample are observed with HNCO fractional abundances ranging from 10$^{-9}$ to 10$^{-8}$, but
for two sources, namely NGC\,5194 and M\,82, showing remarkable low HNCO abundances.
For these sources we derived fractional abundance below the average derived from the rest of galaxies in our sample by a
factor of $\sim3$ and $\sim20$, respectively.

%For the sake of completeness, Fig.~\ref{fig.CSHNCO} includes the CS and C$^{34}$S abundances and the HNCO upper limit abundance based
%on previous observations towards the central position of M\,82 \citep{Martin06b}.
%These abundances are relative to the H$_2$ derived from $^{13}$CO \citep{Mauers03} so they might be overestimated in a factor $\sim2-3$.
%Even on its central position, the HNCO abundance in M\,82 is significantly lower than the average value in the rest of observed galaxies.

%In the following discussion we will focus mainly in the result from the $^{34}$CS-to-HNCO abundance comparison.
We do not observe major differences in the derived abundances among the AGNs and SBGs in our sample.
This homogeneity suggests that the type of nuclear activity does not play a major role at large scales in the production/destruction of these species
in the bulk of the molecular gas.
%Indeed the three AGNs in this work show all similar abundances, where the small differences are likely due to the contribution of the starburst
%components around the AGNs.
This effect is illustrated by the two positions observed in NGC\,1068, where the HNCO and $^{34}$CS abundances decrease by less than a factor of two
from the nuclear AGN to the SB ring.
Of course we have to bear in mind that the two regions are not fully resolved by the beam of the IRAM 30\,m telescope at these frequencies.

If, as suggested by the observations in the Galactic Center region \citep{Martin08}, the abundance of these species is mostly driven by the presence of
shocks and strong UV radiation fields in the ISM, then the observed differences are likely linked to star formation activity in these galaxies.
Significant differences are observed within the galaxies IC\,342 and Maffei\,2.
In IC\,342, the HNCO/C$^{18}$O ratio increases towards the position in the minispiral and decreases in the central region where ISM is illuminated by
the UV radiation from the central clusters. These chemical differences, smoothed to the $10''-20''$ single-dish beam size, are much more contrasted
in the high-resolution interferometric maps by \citep{Meier05}.
Although not spatially, we resolve the nuclear region of Maffei\,2 by the two velocity components.
As seen from the high-resolution maps of $^{13}$CO \citep{Meier08} the lower velocity component ($-100$\,\kms) is associated to the nuclear and northeastern molecular arm,
while the higher velocity component (0\,\kms) will be mostly associated to the nuclear and southwest molecular arm.
We clearly see a difference in the chemistry of both components with an increase of the HNCO abundance towards the southwest molecular arm of a factor
of $\sim2$ as well as a decrease of the CS abundance of a factor of $3-4$.
Overall, the HNCO/CS ratio increases by almost an order of magnitude.

We can compare the ratios measured in galaxies 
with the abundances derived in the central region of the Milky Way \citep{Martin08}.
The vast majority of galaxies show HNCO abundances intermediate between those found in the shock dominated giant molecular clouds (GMCs) 
and the Galactic PDR prototypes (see dashed lines in Fig~\ref{fig.CSHNCO}).
Two sources in our sample, namely M\,83 and IC\,342$^*$ clearly fall in the range of abundances found in GMCs.
On the contrary, M\,82 represents the other extreme where HNCO emission is utterly vanished.
The upper limit to the detection derived after a deep 5.5 hours integration on the HNCO $5-4$ 
is an order of magnitude lower than the previous limit to the detection in the central position \citep{Nguyen91}, 
but still HNCO remains undetected.
Interferometric maps with the BIMA array shows no emission anywhere in the nuclear region of M\,82 \citep{Turner08}.
This non-detection of HNCO supports the idea that M\,82 is highly pervaded by a strong UV photodissociating radiation
% as traced by HCO, CO$^+$ and HOC$^+$ \citep{Burillo02,Fuente05,Fuente06}, but is highly 
and depleted of dense molecular as also suggested by the low abundance of CH$_3$OH and NH$_3$ \citep{Mauers93,Takano02,Martin06a}.
We would expect to observe a higher abundance of CS in this NE position where photodissociation is claimed to be dominant \citep{Burillo02}.
However, we observe just a slightly higher CS abundance than in the whole sample, and not even that if we consider the isotopologue C$^{34}$S.
Altogether with M\,82, the upper limit to the HNCO abundance derived for NGC5194 are similar to those derived in the Galactic Center PDRs.
Most of the molecular studies of NGC5194 so far are restricted to the emission of CO and HCN \citep[][and references therein]{Kohno96}.
The results from this work point out the possible similarities between the chemistries of NGC\,5194 and M\,82, both with very low
abundances of HNCO as compared with the rest of galaxies in this sample.
This result agrees with the idea of NGC\,5194 being in a post starburst stage after the star formation event triggered by its interaction
with NGC\,5195 \citep{Greenawalt98}.
Additionally, the HNCO and CS abundances in NGC\,253 and NGC\,1068$^*$, a factor of 3 below the average, as well as the significantly high C$^{34}$S abundance,
could be explained by photodissociation.
This would suggest that the UV radiation from the evolved SB plays a significant role in the chemistry
of the nuclear ISM in these galaxies.
Therefore NGC\,253 would contain a significant PDR molecular component, similar to what is observed in M\,82.
However, the amounts of dense molecular gas shielded from the UV radiation is still one order of magnitude larger than that observed in M\,82.
%We also notice that Arp\,220, the extreme star forming galaxy in our sample, shows the higher abundances of both CS and HNCO in the sample.

Our observations of HNCO confirm that this molecule is one of the best tracers to study the evolutionary state of nuclear starbursts
in galaxies because of the high contrast shown between the UV pervaded evolved starbursts and the dense and well shielded molecular material
in the early stages of star formation.
This molecule shows a higher abundance variation in galaxies than other dense molecular gas tracer such as methanol \citep{Hutte97,Martin06a}.
On the other hand, CS does not show a significant variation among the observed sources.
%be strongly affected by the UV radiation as observed in the Galactic center molecular clouds \citep{Martin08}.
This might be due to the fact of CS is not only enhanced in regions with strong UV radiation, but its emission also increases towards
the densest molecular cores due to its high critical density.
It is interesting to note that the similar FUV field derived by \citet{Kramer05} towards the nucleus of M\,83 and NGC\,5194 might
explain the different abundances observed in our sample mostly as a consequence of the dense molecular fuel depletion 
and the molecular cloud structure in their nuclei, and
not so much by the differences in the photodissociation fields affecting the nuclear ISM.
However, the CS variations in the Galactic center clouds are more moderate \citep[a factor of $\sim2-3$,][]{Martin08} 
than those of HNCO, and therefore it is likely to be averaged out over the central several hundred parsecs covered by the single dish beam.

\subsection{The evolution of nuclear starbursts}

The highly contrasted variation of HNCO abundances over the galaxy sample in this work
allows us to trace the dense molecular gas available for 
fueling the nuclear starburst of these galaxies.
This is equivalent to defining a chemical timeline in the evolution of the starburst event that could be read from right to left hand side
in Fig.~\ref{fig.CSHNCO}.
It is important to note that the time scales for the chemical processess due to shocks and photodissociation by UV radiation
\citep[$10^4-10^5$\,yr,][]{Bergin98}
are extremely short as compared with the typical time scale of the starburst phenomenon
\citep[$10^7-10^8$\,yr,][]{Coziol96}.

In this scenario, at the onset of the starburst, large amounts of dense molecular material are driven and compressed into the nuclear region of the galaxy
unleashing the star formation.
High abundances of HNCO are injected into the gas-phase within these clouds due to shocks, likely as a consequence of the cloud-cloud collisions, as
observed in the nuclear spiral arms of IC\,342 \citep{Meier05}.
From our galaxy sample, M\,83 is shown to have the highest abundances of HNCO, and similar to those observed in the Galactic center GMCs.
Subsequently, as soon as the first newly massive star clusters are formed, UV illumination begins to come on stage, photodissociating 
the material surrounding them. The HNCO abundance in these regions will be quickly dissociated where the CS molecule will be formed.
A sort of average equilibrium will be reached between the PDR and the dense gas contribution as long as the infall of molecular fuel
keeps feeding the starburst.
This is the status in which most of the starburst galaxies in our sample are observed.
The AGN dominated galaxies we have observed also show this average CS/HNCO abundances similar to the intermediate state starbursts,
suggesting the presence of a significant UV radiation field.
Once the galaxy gets closer to the end the short few 10\,Myr starburst period, the fuel supply is cut off or stopped by superwinds.
From our comparative study, this is apparently the point at which NGC\,253 stands, where photodissociation becomes more important and the fuel
reservoir is being consumed by the last gasp of star formation.
Finally, as we observe in M\,82 and NGC\,5194, most of the molecular gas is pervaded by the UV radiation field with barely no dense UV shielded
molecular material left.
The galaxy remains then waiting for a likely next burst of star formation \citep{Coziol96}.
Indeed, chemical enrichment in M\,82 points out to recursive starburst periods \citep{Origlia04}.

\section{Conclusions}
%In this paper we have studied the variation of the abundances of CS and HNCO in a sample of extragalactic chemical prototypes.
%As the main conclusions of this comparative study we point out the following:

From our study of the CS and HNCO abundances in a selected sample of extragalactic sources,
we have found differences of nearly two orders of magnitude in the HNCO/CS abundance ratios among the starburst galaxies.
These differences are in agreement with the results from the observations of Galactic center clouds dominated by various heating mechanisms
suggesting that the observed HNCO abundances could be related to the evolutionary stage of their nuclear starburst.
Shocks affecting the massive dense molecular clouds in the early stages and strong UV fields in the evolved stages
dominate the HNCO abundance in galaxies.
We do not find significant differences in the molecular abundances
of HNCO in galaxies with different type of nuclear activity (i.e. AGN or SB) which implies this species are not affected differently by these processes
at the scales observed by single dish beam telescopes.

As derived from the HNCO abundances we can point out prototypes of these nuclear starburst evolutionary stages.
M\,83 and the nuclear spiral arms in IC\,342 would be representative of the early stage, while M\,82 and NGC\,5194 nuclei would be in the later
stages of evolution.
NGC\,253, claimed to be a prototype of early starburst, shows abundances below the average suggesting an important influence of photodissociation
in the heating of its ISM, and therefore chemically closer to M\,82 than to M\,83.

The $^{32}$S/$^{34}$S isotopic ratio, derived from the C$^{32}$S/C$^{34}$S abundance ratio, is found to be $\sim10$, and thus similar for all
the observed sources.
This ratio, found to be well below the value of $\sim24$ in the Galactic disk, supports the idea of
the $^{34}$S isotope being enriched due to massive star formation events in the nuclear region of galaxies.

The CS emission is suggested to be a good tracer of the molecular gas affected by photodissociation in the Galactic center \citep{Martin08}.
We do not observe a significant contrast in our sample.
Though its abundance is enhanced in UV radiated regions, its emission is also favored in high density molecular
clouds. Thus the overall emission over the central few hundred parsecs of galactic nuclei is averaged out, and no significant trends
in the CS abundances is found in the observed galaxies.

\acknowledgments{
This work has been partially supported by the Spanish Ministerio de Educaci\'on y Ciencia under projects
ESP 2004-00665 and ESP2007-65812-C02-01, 
by the Spanish Ministerio de Ciencia e Innovaci\'on under project AYA2005-07516-C02-02
and by the ``Comunidad de Madrid'' Government under PRICIT project S-0505/ESP-0237 (ASTROCAM).
}
 
{\it Facilities:} \facility{IRAM 30m}

\clearpage

\clearpage

\clearpage

\begin{table}
\begin{center}
\caption{Observational parameters}
\label{tab:sourceandlines}
\begin{tabular}{l c c c}
\tableline
\tableline
\multicolumn{4}{c}{SOURCE SAMPLE} \\
Source                      &  $\alpha_{J2000}$    & $\delta_{J2000}$  &   $D_{\rm L/A}$\,(Mpc$-$pc/$''$)~\tablenotemark{a}  \\
\tableline
{\it NGC\,253}                &  00:47:33.3        & -25:17:23         &   $2.5-12$   \\
Maffei\,2                     &  02:41:55.2        &  59:36:11         &   $5.0-24$   \\
NGC\,1068~\tablenotemark{b}   &  02:42:40.9        & -00:00:46         &   $16.7-81$  \\
IC\,342~\tablenotemark{c}     &  03:46:48.6        &  68:05:46         &   $3.7-18$   \\
M\,82~\tablenotemark{d}       &  09:55:51.9        &  69:40:47         &   $3.5-17$   \\
{\it NGC\,4945}               &  13:05:27.2        & -49:28:05         &   $7.8-38$   \\
NGC\,5194 (M\,51)             &  13:29:52.7        &  47:11:43         &   $9.7-47$   \\
M\,83                         &  13:37:00.9        & -29:51:56         &   $3.7-18$   \\
Arp\,220                      &  15:34:57.1        &  23:30:12         &   $73.0-354$  \\
NGC\,6946                     &  20:34:52.4        &  60:09:14         &   $5.5-27$   \\
NGC\,7469                     &  23:03:15.6        &  08:52:26         &   $66-320$  \\
\tableline
\multicolumn{4}{c}{MOLECULAR LINES} \\
Molecule                             &  Frequency  (GHz)             &   $\Theta_{\rm beam}\, ('')$~\tablenotemark{e} &  $\rm B_{eff}$~\tablenotemark{f}   \\
C$^{18}$O $J=1-0$                    &   109.782                     &     22                                         &  0.75   \\
C$^{18}$O $J=2-1$                    &   219.560                     &     11                                         &  0.55   \\
CS $J=5-4$                           &   244.935                     &     10                                         &  0.49   \\
C$^{34}$S $J=2-1$~ \tablenotemark{g} &    96.412                     &     25                                         &  0.76   \\
C$^{34}$S $J=3-2$                    &   144.617                     &     17                                         &  0.69   \\
HNCO $5_{0,5}-4_{0,4}$               &   109.905                     &     22                                         &  0.75   \\
HNCO $6_{1,6}-5_{1,5}$               &   131.885                     &     19                                         &  0.71   \\
HNCO $10_{0,10}-9_{0,9}$             &   219.798                     &     11                                         &  0.55   \\ 
\tableline
\end{tabular}
\tablecomments{The data for sources in {\it italics} were taken from previously published data.}
\tablenotetext{a}{Distances as derived by \citet{Baan08} and linear scales given in parsecs per arcsecond at those distances.}
\tablenotetext{b}{Central and $(0'',-16'')$ offset (NGC\,1068$^*$ in the text) positions were observed.}
\tablenotetext{c}{Central and $(5'',15'')$ offset (IC\,342$^*$ in the text) positions were observed.}
\tablenotetext{d}{Central position was not observed, but the NE molecular lobe at the approximate offset $(13'',7.5'')$ position,
referred to as M\,82$^*$ in the text.}
\tablenotetext{e}{All data are from the IRAM 30\,m telescope, but for those of NGC\,4945 from SEST \citep{Wang04}.
In this case, the beam is about twice the given size.}
\tablenotetext{f}{Beam efficiency of the IRAM 30\,m at the observed frequencies}
\tablenotetext{g}{This transition is only observed towards Arp\,220.}
\end{center}
\end{table}

\begin{table}
\begin{center}
\tiny
\caption{ Parameters derived from the Gaussian profiles fitted to the observations.\label{tab.gaussfit}}
%\scriptsize
\tiny
\begin{tabular}{l c c c c c}
\tableline
\tableline
Transition             &  $\int{T_{\rm mb}{\rm d}v}$  & $v_{\rm LSR}$   &  $\Delta v_{1/2}$             & $T_{\rm MB}$ & $rms$\tablenotemark{a}\\
                       &  (K\,km\,s$^{-1}$)           & (km\,s$^{-1}$)  &  (km\,s$^{-1}$)               & (mK)         & (mK)                  \\
\tableline
\multicolumn{5}{c}{\bf Maffei\,2}  \\
C$^{18}$O $1-0$        &         $1.2\pm0.2$          &      $-102\pm4$ &       $63\pm7$  &    18.2 &  1.9 \\   %6.2mK/2.72kms
                       &         $3.7\pm0.3$          &         $4\pm4$ &      $120\pm10$ &    29.0 &      \\
C$^{18}$O $2-1$        &         $2.0\pm0.2$          &  $-102.2\pm1.4$ &      $66\pm3$   &    28.0 &  1.0 \\   %2.5mK/5.4kms
                       &         $8.0\pm0.2$          &       $-14\pm2$ &     $151\pm4$   &    49.8 &      \\
CS $5-4$               &         $1.1\pm0.2$          &       $-93\pm4$ &       $63\pm9$  &    16.3 &  1.4 \\   %3.6mK/4.9kms
                       &         $1.7\pm0.3$          &        $2\pm10$ &      $110\pm20$ &    14.3 &      \\
C$^{34}$S $3-2$        &         $0.41\pm0.08$        &       $-100$    &        65       &     6.0 &  1.4 \\   %2.7mK/8.3kms
                       &         $0.37\pm0.12$        &               0 &       110       &     3.2 &      \\
HNCO $5-4$             &         $0.38\pm0.10$        &        $-100$   &        62       &     5.8 &  2.1 \\   %6.9mK/2.7kms
                       &         $1.42\pm0.13$        &        $10\pm3$ &      $66\pm7$   &    20.1 &      \\
HNCO $6-5$             &         $<0.3$               &                 &        65       &  $<5.0$ &  2.5 \\   %4.5mK/9.0kms
                       &         $1.67\pm0.17$        &         $4\pm3$ &      $66\pm8$   &    23.9 &      \\
HNCO $10-9$            &         $0.13\pm0.06$        &        $-100$   &       65        &     1.8 &  1.0 \\   %2.3mK/5.4kms
                       &         $0.39\pm0.06$        &        $5\pm6$  &      $62\pm10$  &     5.9 &      \\
\multicolumn{5}{c}{\bf NGC\,1068}  \\
C$^{18}$O $1-0$        &  $4.45\pm0.12$               &   $1102\pm4$    &   $264\pm8 $    &    15.8 & 0.8 \\  %2.6mK/2.7kms
C$^{18}$O $2-1$        &  $ 2.0\pm0.2 $               &   $1067\pm12$   &   $210\pm30$    &     9.0 & 1.9 \\  %4.4mK/5.4kms
CS $5-4$               &  $ 3.3\pm0.3 $               &   $1125\pm9$    &   $180\pm20$    &    17.6 & 3.0 \\  %7.3mK/4.9kms
C$^{34}$S $3-2$        &  $0.53\pm0.12$               &   $1070\pm30$   &   $230\pm60$    &     2.2 & 1.0 \\  %1.9mK/8.3kms
HNCO $5-4$             &   $0.86\pm0.09$              &   $1086\pm11$   &   $210\pm20$    &     3.9 & 0.8 \\  %2.6mK/2.7kms
HNCO $6-5$             &  $0.56\pm0.07$               &   $1114\pm13$   &   $210\pm30$    &     2.5 & 0.6 \\  %1.1mK/9.1kms
HNCO $10-9$            &   $<0.4$                     &                 &     210         &  $<2.1$ & 1.9 \\  %4.4mK/5.4kms
\multicolumn{5}{c}{\bf NGC\,1068 ($0'',-16''$)}  \\
C$^{18}$O $1-0$        &   $3.39\pm0.14$              &   $1130\pm4$    &   $213\pm10$    &    15.0 & 1.1 \\  %3.6mK/2.7kms
C$^{34}$S $3-2$        &   $<0.2$                     &                 &     170         &  $<1.4$ & 1.2 \\  %2.2mK/8.2kms
HNCO $5-4$             &   $0.41\pm0.11$              &   $1160\pm20$   &   $170\pm40$    &     2.3 & 1.1 \\  %3.6mK/2.7kms
HNCO $10-9$            &   $<0.2$                     &                 &     170         &  $<1.0$ & 0.8 \\  %1.5mK/9.1kms WEIRD WIDE FEATURE
\multicolumn{5}{c}{\bf IC\,342 ($0'',0'')$}  \\
C$^{18}$O $1-0$        &         $4.80\pm0.11$        &    $25.2\pm0.6$ &    $50.1\pm1.2$ &    89.9 & 1.7 \\  %5.8mK/2.7kms
C$^{18}$O $2-1$        &         $8.10\pm0.10$        &    $29.9\pm0.3$ &    $46.3\pm0.6$ &   164.1 & 1.1 \\  %2.7mK/5.4kms
CS $5-4$               &         $2.15\pm0.11$        &    $37.7\pm1.2$ &    $50\pm3$     &    40.3 & 1.8 \\  %4.5mK/4.9kms
C$^{34}$S $3-2$        &         $0.49\pm0.12$        &      $26\pm7$   &    $54\pm12$    &     8.4 & 1.5 \\  %2.8mK/8.3kms
HNCO $5-4$             &         $1.81\pm0.10$        &   $30.5\pm1.4$  &    $52\pm3$     &    33.0 & 1.7 \\  %5.8mK/2.7kms
HNCO $6-5$             &         $1.87\pm0.11$        &   $35.0\pm1.6$  &    $53\pm3$     &    33.0 & 1.9 \\  %3.4mK/9.1kms
HNCO $10-9$            &         $0.31\pm0.06$        &       $24\pm3$  &    $34\pm8$     &     8.5 & 1.1 \\  %2.7mK/5.4kms
\multicolumn{5}{c}{\bf IC\,342 ($5'',15'')$}  \\
C$^{18}$O $1-0$        &        $3.45\pm0.10$         &    $43.9\pm0.6$ &  $41.1\pm1.4$   &    78.8 & 2.0 \\  %6.7mK/2.7kms
CS $5-4$               &        $0.42\pm0.10$         &    $49$         &  $36$           &    11.0 & 2.6 \\  %6.5mK/4.9kms
C$^{34}$S $3-2$        &        $0.32\pm0.04$         &    $49\pm2$     &  $36\pm6$       &     8.3 & 1.0 \\  %1.9mK/8.3kms
HNCO $5-4$             &        $1.39\pm0.10$         &    $44.9\pm1.2$ &  $36\pm3$       &    36.2 & 2.0 \\  %6.7mK/2.7kms
HNCO $6-5$             &        $0.81\pm0.10$         &    $41\pm2$     &  $34\pm5$       &    22.7 & 2.0 \\  %3.7mK/9.1kms
\multicolumn{5}{c}{\bf M\,82}  \\
C$^{18}$O $1-0$        &   $1.55\pm0.07$              & $235\pm2$       &    $87\pm5$     & 16.8    & 0.7 \\  % 2.4mk/2.7kms
                       &   $4.41\pm0.06$              & 310             &    80           & 51.9    &     \\  
C$^{18}$O $2-1$        &   $22.79\pm0.15$             & $309.5\pm0.3$   &    $81.9\pm0.6$ & 261.5   & 2.0 \\  % 4.7mk/5.4kms
CS $5-4$               &   $6.4\pm0.2$                & $313.0\pm1.0$   &    $73\pm2$     & 82.6    & 2.5 \\  % 6.1mK/4.9kms 
C$^{34}$S $3-2$        &   $0.58\pm0.08$              & $302\pm6$       &    $91\pm13$    &  6.0    & 1.0 \\  % 1.95mK/8.3kms
HNCO $5-4$             &   $<0.10$                    &                 &    80           & $<1.3$  & 0.7 \\  % 2.37mk/2.7kms 
HNCO $6-5$             &   $<0.17$                    &                 &    80           & $<2.3$  & 1.3 \\  % 2.31mK/9.09kms
HNCO $10-9$            &   $<0.3$                     &                 &    80           & $<3.7$  & 2.0 \\  % 4.75mk/5.4kms 
\multicolumn{5}{c}{\bf NGC\,5194}  \\
C$^{18}$O $1-0$        &   $0.80\pm0.08$              &   $418.2\pm1.4$ &   $39\pm3$      &    19.1 & 0.6 \\ %2.1mK/2.7kms
                       &   $0.86\pm0.09$              &   $490\pm5$     &   $80\pm8$      &    10.1 &     \\
C$^{34}$S $3-2$        &   $<0.2$                     &                 &   100           &  $<2.2$ & 1.4 \\ %2.6mK/8.3kms 
HNCO $5-4$             &   $<0.11$                    &                 &   100           &  $<1.0$ & 0.6 \\ %2.1mK/2.7kms
\multicolumn{5}{c}{\bf M\,83}  \\
C$^{18}$O $1-0$        &   $2.19\pm0.14$              &   $499\pm2$     &  $66\pm5$       &    30.9 & 1.9 \\ % 6.4mK/2.7kms
CS $5-4$               &   $2.3\pm0.3$                &   $504\pm7$     &  $82.4\pm1.1$   &    26.5 & 0.5 \\ % 1.3mK/4.9kms
C$^{34}$S $3-2$        &   $0.40\pm0.13$              &   $501\pm6$     &  $34\pm13$      &    11.1 & 2.7 \\ % 5.1mK/8.3kms NOISY WIDE FEATURE
HNCO $5-4$             &   $0.70\pm0.15$              &   $510\pm10$    &  $90\pm19$      &     7.7 & 1.9 \\ % 6.4mK/2.7kms
HNCO $6-5$             &   $0.38\pm0.07$              &   $515\pm7$     &  $67\pm12$      &     5.4 & 1.2 \\ % 2.1mK/9.1kms
\multicolumn{5}{c}{\bf Arp\,220}  \\
C$^{18}$O $1-0$        &   $2.7\pm0.3$                &   $5445$        &   $450\pm40$    &     5.7 & 0.9 \\  % 3.0mK/2.7kms
CS $5-4$               &   $11.3\pm0.2$               &   $5446\pm5$    &   $481\pm11$    &    22.0 & 2.7 \\  % 3.9mK/14.7kms
C$^{34}$S $2-1$        &   $0.9\pm0.2$                &   $5460\pm30$   &   $240\pm50$    &     3.4 & 1.4 \\  % 4.3mK/3.1kms
HNCO $5-4$             &   $1.1\pm0.2$                &   $5445$        &   $450$         &     2.3 & 0.9 \\  % 3.0mK/2.7kms
\multicolumn{5}{c}{\bf NGC\,6946}  \\
C$^{18}$O $1-0$        &   $1.36\pm0.10$              &   $-5.0\pm1.4$  &  $ 54\pm3$      &    23.8 & 0.7 \\ %2.2mK/2.7kms
                       &   $3.41\pm0.11$              &   $84.7\pm1.7$  &  $103\pm4$      &    31.2 &     \\
C$^{34}$S $3-2$        &   $0.37\pm0.07$              &   $71\pm11$     &  $ 90\pm20$     &     3.8 & 0.7 \\ %1.3mK/8.2kms  VER ESTE AJUSTE
HNCO $5-4$             &   $0.55\pm0.10$              &   $62\pm8$      &  $87\pm19$      &     5.9 & 0.7 \\ %2.2mK/2.7kms
HNCO $6-5$             &   $0.24\pm0.06$              &   $54\pm7$      &  $53\pm14$      &     4.1 & 1.0 \\ %1.9mK/9.1kms
\multicolumn{5}{c}{\bf NGC\,7469}  \\
C$^{18}$O $1-0$        &   $0.61\pm0.05$              &   $4910\pm10$   &  $250\pm20$     &     2.3 & 0.4 \\ %0.76mK/10.9kms
C$^{34}$S $3-2$        &   $<0.13$                    &                 &  150            &  $<0.9$ & 0.6 \\ %1.24mK/8.3kms
HNCO $5-4$             &   $0.16\pm0.04$              &   $4890\pm20$   &  $150\pm40$     &     1.0 & 0.4 \\ %0.76mK/10.9kms
\tableline
\tablecomments{Parameters without error indicate fixed for the Gaussian fitting. Upper limits correspond to $3\sigma$ limits.}
\tablenotetext{a}{$1\sigma$ rms in 30\,\kms channels.}
\end{tabular}
\end{center}
\end{table}

 \begin{table}
\setlength{\tabcolsep}{0.03in}
\begin{center}
\caption{Derived column densities and rotational temperatures for each galaxy.\label{tab.coldens}}
\begin{tabular}{lccccccc}
\tableline
\tableline
Source                      &         \multicolumn{2}{c}{C$^{18}$O}            &    CS                               &    C$^{34}$S                   &        \multicolumn{2}{c}{HNCO}      \\
                            &($\times10^{15}\rm{cm}^{-2}$) & (K)               &($\times10^{13}\rm{cm}^{-2}$)        &($\times10^{12}\rm{cm}^{-2}$)   &($\times10^{13}\rm{cm}^{-2}$) & (K)   \\
\tableline
NGC\,253 \tablenotemark{a}  &       $100\pm20  $           & $5.6\pm0.4$       &  $21\pm4   $                        &     $19\pm4    $               &    $  22\pm9    $            & $18\pm4$        \\
Maffei\,2 \tablenotemark{b} &       $5.7\pm1.5 $           & $5.6\pm0.6$       &  $2.4\pm0.7$                        &     $5.8\pm1.1 $               &    $  2.5\pm1.2 $            & $12.2\pm1.9$    \\
                            &       $17\pm2    $           & $6.6\pm0.3$       &  $3.5\pm0.9$                        &     $5.2\pm1.0 $               &    $  10\pm2    $            & $11.5\pm0.6$    \\
NGC\,1068                   &       $24\pm2    $           & $3.3\pm0.1$       &  $7.0\pm1.5$                        &     $7.3\pm1.2 $               &    $  10\pm5    $            & $ 5.8\pm0.8$    \\
NGC\,1068\tablenotemark{*}  &       $18\pm4    $           &   ...             &   ...                               &     $<3.4      $               &    $  2.9\pm1.0 $            &    ...          \\
IC\,342                     &       $22\pm1    $           & $5.7\pm0.1$       &  $4.6\pm0.1$                        &     $6.9\pm1.3 $               &    $  12\pm1    $            & $10.2\pm0.4$    \\
IC\,342\tablenotemark{*}    &       $18\pm4    $           &   ...             &  $0.9\pm0.2$                        &     $4.4\pm1.0 $               &    $  19\pm1    $            & $ 5.3\pm0.6$    \\
M\,82\tablenotemark{*}      &       $28\pm1    $           & $14.5\pm0.3$      &  $14\pm4   $                        &     $8.1\pm1.9 $               &    $  <0.7      $            &   ...           \\
NGC\,4945\tablenotemark{c}  &       $160\pm10  $           & $6.5\pm0.2$       &  $58\pm12  $                        &     $64\pm13   $               &    $  50\pm16   $            &    ...          \\
NGC\,5194                   &       $8.6\pm1.8 $           &   ...             &  $<0.7     $\tablenotemark{d}       &     $<3.0      $               &    $  <0.7      $            &    ...          \\
M\,83                       &       $11\pm2    $           &   ...             &  $4.9\pm1.2$                        &     $5.6\pm1.1 $               &    $  10\pm5    $            & $ 5.0\pm1.1$    \\
Arp\,220                    &       $14\pm3    $           &   ...             &  $24\pm5   $                        &     $26\pm5    $               &    $  7\pm3     $            &    ...          \\
NGC\,6946                   &       $25\pm5    $           &   ...             &  $6.0\pm0.8$\tablenotemark{d}       &     $5.2\pm1.0 $               &    $  12\pm6    $            & $ 4.2\pm0.9$    \\
NGC\,7469                   &       $3.2\pm0.7 $           &   ...             &   ...                               &     $<1.9      $               &    $  1.1\pm0.4 $            &    ...          \\
\tableline
\end{tabular}
%% Any table notes must follow the \end{tabular} command.
%\tablenotetext{a}{$T_{\rm ex}$ value fixed to 10\,K.}
\tablenotetext{*}{\,Indicates offset positions.}
\tablenotetext{a}{\,Derived using data from \citet{Harrison99,Martin05,Martin06b}.}
\tablenotetext{b}{\,Abundances derived for each of the two velocity components labeled as Maffei\,2a and Maffei\,2b in Fig.~\ref{fig.CSHNCO}.}
\tablenotetext{c}{\,Derived using data from \citet{Wang04}.}
\tablenotetext{d}{\,Derived using data from \citet{Mauers89}.}
\end{center}
\end{table}


\begin{thebibliography}{}
\bibitem[Aalto et al.(2002)]{Aalto02} Aalto, S., Polatidis, A.~G., H{\"u}ttemeister, S., \& Curran, S.~J.\ 2002, \aap, 381, 783 
\bibitem[Aalto et al.(2007)]{Aalto07} Aalto, S., Spaans, M., Wiedner, M.C., \& H\"uttemeister, S. 2007, \aap, 464, 193
\bibitem[Antonucci \& Miller(1985)]{Anto85} Antonucci, R.~R.~J., \& Miller, J.~S.\ 1985, \apj, 297, 621 
\bibitem[Baan \& Haschick(1995)]{Baan95} Baan, W.~A., \& Haschick, A.~D.\ 1995, \apj, 454, 745 
\bibitem[Baan et al.(2008)]{Baan08} Baan, W.~A., Henkel, C., Loenen, A.~F., Baudry, A., \& Wiklind, T.\ 2008, \aap, 477, 747 
\bibitem[Bergin et al.(1998)]{Bergin98} Bergin, E.~A., Neufeld, D.~A., \& Melnick, G.~J.\ 1998, \apj, 499, 777 
\bibitem[Boker et al.(1997)]{Boker97} Boker, T., Forster-Schreiber, N.~M., \& Genzel, R.\ 1997, \aj, 114, 1883 
\bibitem[Braatz et al.(1997)]{Braatz97} Braatz, J.~A., Wilson, A.~S., \& Henkel, C.\ 1997, \apjs, 110, 321 
%%\bibitem[Charnley et al.(1995)]{Charnley95} Charnley, S.B., Kress, M.E., Tielens, A.G.G.M., \& Millar, T.J. 1995, ApJ, 448, 232
\bibitem[Chin et al.(1996)]{Chin96} Chin, Y.-N., Henkel, C., Whiteoak, J.B., Langer, N., \& Churchwell, E.B. 1996, \aap, 305, 960
%%\bibitem[de Vicente et al. (2000)]{deVicente00} de Vicente, P., Mart\'{\i}n-Pintado, J., Neri, R., \& Colom, P. 2000, \aap, 361, 1058
%%\bibitem[Dahmen et al.(1997)]{Dahmen97} Dahmen, G., H\"uttemeister, S., Wilson, T.L., et al. 1997, A\&AS, 126, 197
%%\bibitem[Drdla et al.(1989)]{Drdla89} Drdla, K., Knapp, G. R., \& van Dishoeck, E. F. 1989, ApJ, 345, 815
\bibitem[Coziol(1996)]{Coziol96} Coziol, R.\ 1996, \aap, 309, 345 
\bibitem[Dopita et al.(2005)]{Dopita05} Dopita, M.~A., et al.\ 2005, \apj, 619, 755 
%%\bibitem[Eisenhauer et al.(2005)]{Eisen05} Eisenhauer, F., Genzel, R., Alexander, T. et al. 2005, ApJ, 628, 246
%%\bibitem[Figer et al.(1999)]{Figer99} Figer, D.F.; McLean, I.S., \& Morris, M. 1999, ApJ, 514, 202
%%\bibitem[Frerking et al.(1982)]{Frerking82} Frerking, M. A., Langer, W. D., \& Wilson, R. W. 1982, ApJ, 262, 590
\bibitem[Fuente et al.(2005)]{Fuente05} Fuente, A., Garc\'{\i}a-Burillo, S., Gerin, M., Teyssier, D., Usero, A., Rizzo, J. R., \& de Vicente, P. 2005, \apj, 619, L155  
\bibitem[Fuente et al.(2006)]{Fuente06} Fuente, A., Garc{\'{\i}}a-Burillo, S., Gerin, M., Rizzo, J.~R., Usero, A., Teyssier, D., Roueff, E., \& Le Bourlot, J.\ 2006, \apjl, 641, L105 
%\bibitem[Gao \& Solomon(2004a)]{Gao04a} Gao, Y., \& Solomon, P.~M.\ 2004a, \apjs, 152, 63 
%\bibitem[Gao \& Solomon(2004b)]{Gao04b} Gao, Y., \& Solomon, P.~M.\ 2004b, \apj, 606, 271 
\bibitem[Garc{\'{\i}}a-Burillo et al.(2002)]{Burillo02} Garc{\'{\i}}a-Burillo, S., Mart{\'{\i}}n-Pintado, J., Fuente, A., Usero, A., \& Neri, R.\ 2002, \apjl, 575, L55 
%%\bibitem[Ghez et al.(2005)]{Ghez05} Ghez, A. M., Salim, S., Hornstein, S. D. 2005, ApJ, 620, 744
\bibitem[Goicoechea et al.(2006)]{Goico06} Goicoechea, J. R.; Pety, J.; Gerin, M. et al. 2006, A\&A, 456, 565
%%\bibitem[Goicoechea et al.(2006)]{Goico06} Goicoechea, J. R.; Pety, J.; Gerin, M. et al. 2006, A\&A, 456, 565
%\bibitem[Graci{\'a}-Carpio et al.(2008)]{Gracia08} Graci{\'a}-Carpio, J., Garc{\'{\i}}a-Burillo, S., Planesas, P., Fuente, A., \& Usero, A.\ 2008, \aap, 479, 703 
%%\bibitem[G\"usten et al.(1981)]{Gusten81} G\"usten, R., Walmsley, C.M., Pauls T. 1981, A\&A, 103, 197
%%\bibitem[Hasegawa \& Herbst(1993)]{Hasegawa93} Hasegawa, T. I., \& Herbst, E. 1993, MNRAS, 263, 589
%%\bibitem[Henkel et al.(1997)]{Henkel97} Henkel, C., Jacq, T., Mauersberger, R., Menten, K.M., Steppe, H. 1997, \aap, 188, L1
%%\bibitem[H\"uttemeister et al.(1993)]{Hutte93} H\"uttemeister, S., Wilson, T.L., Bania, T.M., \& Mart\'{\i}n-Pintado, J. 1993, \aap, 280, 255
%%\bibitem[H\"uttemeister et al.(1998)]{Hutte98} H\"uttemeister, S., Dahmen, G., Mauersberger, R., Henkel, C., Wilson, T. L., \& Mart\'{\i}n-Pintado, J. 1998, A\&A, 334, 646
%%\bibitem[Iglesias(1977)]{Iglesias77}Iglesias, E. 1977, ApJ, 218, 697
%%\bibitem[Jansen et al.(1995)]{Jansen95}Jansen, D. J., Spaans, M., Hogerheijde, M. R., \& Van Dishoeck, E. F. 1995, \aap, 303, 541 
%%\bibitem[Jackson et al.(1984)]{Jackson84} Jackson, J.M., Armstrong, J.T., Barrett, A.H. 1984, ApJ, 280, 608
\bibitem[Greenawalt et al.(1998)]{Greenawalt98} Greenawalt, B., Walterbos, R.~A.~M., Thilker, D., \& Hoopes, C.~G.\ 1998, \apj, 506, 135 
\bibitem[Harrison et al.(1999)]{Harrison99} Harrison, A., Henkel, C., \& Russell, A.\ 1999, \mnras, 303, 157 
\bibitem[Helfer et al.(2003)]{Helfer03} Helfer, T.~T., Thornley, M.~D., Regan, M.~W., Wong, T., Sheth, K., Vogel, S.~N., Blitz, L., \& Bock, D.~C.-J.\ 2003, \apjs, 145, 259 
%%\bibitem[Hudson et al.(2001)]{Hudson01}Hudson, R. L., Moore, M. H., \& Gerakines, P. A. 2001, ApJ, 550, 1140
\bibitem[Hurt et al.(1996)]{Hurt96} Hurt, R.~L., Turner, J.~L., \& Ho, P.~T.~P.\ 1996, \apj, 466, 135 
\bibitem[H\"uttemeister et al.(1997)]{Hutte97} H\"uttemeister, S., Mauersberger, R., \& Henkel, C.\ 1997, \aap, 326, 59 
\bibitem[Ishizuki et al.(1990)]{Ishizuki90} Ishizuki, S., Kawabe, R., Ishiguro, M., Okumura, S.~K., \& Morita, K.-I.\ 1990, \nat, 344, 224 
\bibitem[Kramer et al.(2005)]{Kramer05} Kramer, C., Mookerjea, B., Bayet, E., Garcia-Burillo, S., Gerin, M., Israel, F.~P., Stutzki, J., \& Wouterloot, J.~G.~A.\ 2005, \aap, 441, 961 
\bibitem[Kohno et al.(1996)]{Kohno96} Kohno, K., Kawabe, R., Tosaki, T., \& Okumura, S.~K.\ 1996, \apjl, 461, L29 
\bibitem[Kohno et al.(1999)]{Kohno99} Kohno, K., Matsushita, S., Vila-Vilar\'o, B., Okumura, S. K., Shibatsuka, T., Okiura, M., Ishizuki, S., \& Kawabe, R. 2001, ASP Conference Proceedings Vol. 249: ``The Central Kiloparsec of Starbursts and AGN: The La Palma Connection", 672
\bibitem[Kohno(2005)]{Kohno05} Kohno, K. 2005, AIP conference proceedings 783 "The evolution of Starburst: The 331st Wilhelm and Else Heraeus Seminar", 203
%%\bibitem[Kohno et al.(2007)]{Kohno07} Kohno, K., Nakanishi, K., \& Imanishi, M. 2007, proceedings "The Central Engine of Active Galactic Nuclei", ed. L. C. Ho and J.-M. Wang, astro-ph/0704.2818
%%\bibitem[Krabbe et al.(1995)]{Krabbe95} Krabbe, A., Genzel, R., Eckart, A., \& Najarro, F. 1995, ApJ, 447, L95
\bibitem[Krips et al.(2008)]{Krips08} Krips, M., Neri, R., Garc{\'{\i}}a-Burillo, S., Mart{\'{\i}}n, S., Combes, F., Graci{\'a}-Carpio, J., \& Eckart, A.\ 2008, \apj, 677, 262 
%%\bibitem[Lindqvist et al.(1995)]{Lindqvist95} Lindqvist, M., Sandqvist, A., Winnberg, A., Johansson, L.E.B., Nyman, \& L.-A. 1995, A\&AS, 113, 257L
\bibitem[Maiolino et al.(1999)]{Maio99} Maiolino, R., Risaliti, G., \& Salvati, M.\ 1999, \aap, 341, L35 
\bibitem[Marconi et al.(2000)]{Marconi00} Marconi, A., Oliva, E., van der Werf, P.~P., Maiolino, R., Schreier, E.~J., Macchetto, F., \& Moorwood, A.~F.~M.\ 2000, \aap, 357, 24 
%%\bibitem[Mason \& Wilson(2004)]{Mason04} Mason, A.~M., \& Wilson, C.~D.\ 2004, \apj, 612, 860 
\bibitem[Mart{\'{\i}}n et al.(2003)]{Martin03} Mart{\'{\i}}n, S., Mauersberger, R., Mart{\'{\i}}n-Pintado, J., Garc{\'{\i}}a-Burillo, S., \& Henkel, C.\ 2003, \aap, 411, L465 
\bibitem[Mart{\'{\i}}n et al.(2005)]{Martin05} Mart{\'{\i}}n, S., Mart{\'{\i}}n-Pintado, J., Mauersberger, R., Henkel, C., \& Garc{\'{\i}}a-Burillo, S.\ 2005, \apj, 620, 210 
\bibitem[Mart\'{\i}n et al.(2006a)]{Martin06a} Mart\'{\i}n, S., Mart\'{\i}n-Pintado, J., Mauersberger, R. 2006a, \aap, 450 L13
\bibitem[Mart\'{\i}n et al.(2006b)]{Martin06b} Mart\'{\i}n, S., Mauersberger, R., Mart\'{\i}n-Pintado, J., \& Henkel, C., Garc\'{\i}a-Burillo, S. 2006b, \apjs, 164, 450
\bibitem[Mart\'{\i}n et al.(2008)]{Martin08} Mart\'{\i}n, S., Requena-Torres, M.~A., Mart\'{\i}n-Pintado, J., \& Mauersberger, R.\ 2008, \apj, 678, 245
%%\bibitem[Mart\'{\i}n-Pintado et al.(1997)]{Pintado97} Mart\'{\i}n-Pintado, J., de Vicente, P., Fuente, A., \& Planesas, P. 1997, ApJ,482, 45
%%\bibitem[Mart\'{\i}n-Pintado et al.(2001)]{Pintado01} Mart\'{\i}n-Pintado, J., Rizzo, J.R., de Vicente, P., Rodr\'{\i}guez-Fern\'andez, N.J., \& Fuente, A. 2001, ApJ, 548, L65
%%\bibitem[Mart\'{\i}n-Pintado et al.(2007)]{Pintado07} Mart\'{\i}n-Pintado et al. 2007, in preparation 
\bibitem[Mason \& Wilson(2004)]{Mason04} Mason, A.~M., \& Wilson, C.~D.\ 2004, \apj, 612, 860 
%%\bibitem[Mauersberger \& Henkel(1989)]{Mauers89}Mauersberger, R., \& Henkel, C. 1989, A\&A, 223, 79
\bibitem[Mauersberger et al.(1989)]{Mauers89} Mauersberger, R., Henkel, C., Wilson, T.L., \& Harju, J. 1989, A\&A, 226, L5
\bibitem[Mauersberger \& Henkel(1993)]{Mauers93} Mauersberger, R. \& Henkel, C. 1993, Rev. Modern Astron. 6, 69
\bibitem[Mauersberger et al.(2003)]{Mauers03} Mauersberger, R., Henkel, C., Wei\ss, A., Peck, A.B., Hagiwara, Y. 2003, \aap, 403, 561
%%\bibitem[McQuinn et al.(2002)]{McQuinn02} McQuinn, K.B.W., Simon, R.,  Law, C.J. et al. 2002, ApJ, 576, 274
\bibitem[Meier \& Turner(2005)]{Meier05} Meier, D.~S., \& Turner, J.~L.\ 2005, \apj, 618, 259 
\bibitem[Meier et al.(2008)]{Meier08} Meier, D.~S., Turner, J.~L., \& Hurt, R.~L.\ 2008, \apj, 675, 281 
\bibitem[Myers \& Scoville(1987)]{Myers87} Myers, S.~T., \& Scoville, N.~Z.\ 1987, \apjl, 312, L39 
%%\bibitem[Millar et al.(1991)]{Millar91} Millar, T.J., Herbst, E., \& Charnley, S.B. 1991, ApJ, 369, 147
%%\bibitem[Minh et al.(1998)]{Minh98} Minh, Y.C., Haikala, L., Hjalmarson, \AA., \& Irvine, W. M. 1998, ApJ, 498, 261
%%%\bibitem[Minh et al.(2005)]{Minh05} Minh, Y.C., Kim, S.-J., Pak, S., Lee, S., Irvine, W. M., Nyman, L.-\AA. 2005, New Astronomy, 10, 425 
%%\bibitem[Minh \& Irvine(2006)]{Minh06} Minh, Y.C. \& Irvine, W.M. 2006, New Astronomy, 11, 594 
%%\bibitem[Morris \& Serabyn(1996)]{Morris96} Morris, M. \& Serabyn, E. 1996, ARA\&A, 34, 645
%\bibitem[Narayanan et al.(2005)]{Narayanan05} Narayanan, D., Groppi, C.~E., Kulesa, C.~A., \& Walker, C.~K.\ 2005, \apj, 630, 269 
%\bibitem[Narayanan et al.(2007)]{Narayanan07} Narayanan, D., Cox, T.~J., Shirley, Y., Dave, R., Hernquist, L., \& Walker, C.~K.\ 2007, ArXiv e-prints, 711, arXiv:0711.1361 
\bibitem[Nguyen-Q-Rieu et al.(1991)]{Nguyen91} Nguyen-Q-Rieu, Henkel, C., Jackson, J.~M., \& Mauersberger, R.\ 1991, \aap, 241, L33 
%%\bibitem[Novozamsky et al.(2001)]{Novaz01} Novozamsky, J.H., Schutte, W.A., \& Keane, J.V. 2001, A\&A, 379, 588
%%\bibitem[Oka et al.(1998)]{Oka98} Oka, T., Hasegawa, T., Sato, F., Tsuboi, M., Miyazaki, A. 1998, ApJS, 118, 455 
%%\bibitem[Oka et al.(1999)]{Oka99} Oka, T., White, G.J., Hasegawa, T., Sato, F., Tsuboi, M., Miyazaki, A. 1999, ApJ, 515, 249
\bibitem[Origlia et al.(2004)]{Origlia04} Origlia, L., Ranalli, P., Comastri, A., \& Maiolino, R.\ 2004, \apj, 606, 862 
\bibitem[Petitpas \& Wilson(1998)]{Petipas98} Petitpas, G.~R., \& Wilson, C.~D.\ 1998, \apj, 503, 219 
\bibitem[Rekola et al.(2005)]{Rekola05} Rekola, R., Richer, M.~G., McCall, M.~L., Valtonen, M.~J., Kotilainen, J.~K., \& Flynn, C.\ 2005, \mnras, 361, 330 
%%\bibitem[Requena-Torres et al.(2006)]{Requena06} Requena-Torres, M.A., Mart\'{\i}n-Pintado, J., Rodr\'{\i}guez-Franco, A, et al. 2006, A\&A, 455, 971
%%\bibitem[Requena-Torres et al.(2007)]{Requena07} Requena-Torres, Mart\'{\i}n-Pintado, J., Mart\'{\i}n, S., Morris, M. 2007, ApJ Accepted
%%\bibitem[Rodr\'{\i}guez-Fern\'andez et al.(2001)]{Rodriguez01}Rodr\'{\i}guez-Fern\'andez, N.J., Mart\'{\i}n-Pintado, J., \&  de Vicente, P. 2001, \aap, 377, 631
%%\bibitem[Roberge et al.(1991)]{Roberge91} Roberge, W.G., Jones, D., Lepp, S., Dalgarno, A. 1991, ApJS, 77, 287
\bibitem[Sakamoto et al.(1999)]{Saka99} Sakamoto, K., Scoville, N.~Z., Yun, M.~S., Crosas, M., Genzel, R., \& Tacconi, L.~J.\ 1999, \apj, 514, 68 
\bibitem[Sanders et al.(2003)]{Sanders03} Sanders, D.~B., Mazzarella, J.~M., Kim, D.-C., Surace, J.~A., \& Soifer, B.~T.\ 2003, \aj, 126, 1607 
%%\bibitem[Serabyn \& G\"usten(1991)]{Serabyn91} Serabyn, E. \& G\"usten, R. 1991, \aap, 242, 376
\bibitem[Schinnerer et al.(2000)]{Schinne00} Schinnerer, E., Eckart, A., Tacconi, L.~J., Genzel, R., \& Downes, D.\ 2000, \apj, 533, 850 
\bibitem[Schinnerer et al.(2007)]{Schinne07} Schinnerer, E., B{\"o}ker, T., Emsellem, E., \& Downes, D.\ 2007, \aap, 462, L27 
\bibitem[Sch{\"o}ier et al.(2005)]{Schoier05} Sch{\"o}ier, F.~L., van der Tak, F.~F.~S., van Dishoeck, E.~F., \& Black, J.~H.\ 2005, \aap, 432, 369 
\bibitem[Schuster et al.(2007)]{Schuster07} Schuster, K.~F., Kramer, C., Hitschfeld, M., Garcia-Burillo, S., \& Mookerjea, B.\ 2007, \aap, 461, 143 
\bibitem[Scoville \& Young(1983)]{Scoville83} Scoville, N., \& Young, J.~S.\ 1983, \apj, 265, 148 
\bibitem[Scoville et al.(2000)]{Scov00} Scoville N.~Z. et al., 2000, \aj, 119, 991
\bibitem[Scoville et al.(2001)]{Scoville01} Scoville, N.~Z., Polletta, M., Ewald, S., Stolovy, S.~R., Thompson, R., \& Rieke, M.\ 2001, \aj, 122, 3017 
%%\bibitem[Shull \& McKee(1979)]{Shull79} Shull, J. M., \& McKee, C. F. 1979, ApJ, 227
%%\bibitem[Sidoli et al.(2001)]{Sidoli01} Sidoli, L.; Mereghetti, S.; Treves, A.; Parmar, A. N.; Turolla, R.; Favata, F. 2001, A\&aA, 372, 651
\bibitem[Soifer et al.(1989)]{Soifer89} Soifer, B.T., Boehmer, L., Neugebauer, G., \& Sanders, D.B. 1989, AJ, 98, 766
%\bibitem[Solomon et al.(1997)]{Solomon97} Solomon, P.~M., Downes, D., Radford, S.~J.~E., \& Barrett, J.~W.\ 1997, \apj, 478, 144 
\bibitem[Sternberg \& Dalgarno(1995)]{Stern95} Sternberg, A., \& Dalgarno, A. 1995, ApJS, 99, 565
\bibitem[Talbot et al.(1979)]{Talbot79} Talbot, R.~J., Jr., Jensen, E.~B., \& Dufour, R.~J.\ 1979, \apj, 229, 91 
%\bibitem[Takano et al.(1995)]{Takano95} Takano, S., Nakai, N., \& Kawaguchi, K. 1995, PASJ, 47 171
\bibitem[Takano et al.(2002)]{Takano02} Takano, S., Nakai, N., \& Kawaguchi 2002, PASJ, 54, 195
\bibitem[Thronson et al.(1991)]{Thronson91} Thronson, H.~A., Jr., Rubin, H., \& Ksir, A.\ 1991, \mnras, 252, 550 
\bibitem[Turner \& Ho(1983)]{Turner83} Turner, J.~L., \& Ho, P.~T.~P.\ 1983, \apjl, 268, L79 
\bibitem[Turner \& Meier(2008)]{Turner08} Turner, J.~L., \& Meier, D.~S.\ 2008, \apss, 313, 267 
%%%\bibitem[Turner(1996)]{Turner96} Turner B.E. 1996, ApJ, 461, 246
%%\bibitem[Turner et al.(1999)]{Turner99}Turner, B.E., Terzieva, R., \& Herbst, E. 1999, ApJ, 518, 699
%%\bibitem[Usero et al.(2004)]{Usero04} Usero, A., Garc\'{\i}a-Burillo, S., Fuente, A., Mart\'{\i}n-Pintado, J., \& Rodr\'{\i}guez-Fern\'andez, N. J. 2004, \aap, 419, 897
\bibitem[Usero et al.(2006)]{Usero06} Usero, A., Garc{\'{\i}}a-Burillo, S., Mart{\'{\i}}n-Pintado, J., Fuente, A., \& Neri, R.\ 2006, \aap, 448, 457 
\bibitem[van der Tak et al.(2008)]{vdtak08} van der Tak, F.~F.~S., Aalto, S., \& Meijerink, R.\ 2008, \aap, 477, L5 
%%\bibitem[Viti et al.(2002)]{Viti02} Viti, S., Natarajan, S., \& Williams, D.A. 2002, MNRAS, 336, 797
\bibitem[Wang et al.(2004)]{Wang04} Wang, M., Henkel, C., Chin, Y.-N., Whiteoak, J.~B., Hunt Cunningham, M., Mauersberger, R., \& Muders, D.\ 2004, \aap, 422, 883 
\bibitem[Wilson et al.(1991)]{Wilson91} Wilson, A.~S., Helfer, T.~T., Haniff, C.~A., \& Ward, M.~J.\ 1991, \apj, 381, 79 
%\bibitem[Wu et al.(2005)]{Wu05} Wu, J., Evans, N.~J., II, Gao, Y., Solomon, P.~M., Shirley, Y.~L., \& Vanden Bout, P.~A.\ 2005, \apjl, 635, L173 
%%\bibitem[Yusef-Zadeh et al.(1984)]{Yusef84} Yusef-Zadeh, F., Morris, M., \& Chance, D. 1984, \nat, 310, 557
%%\bibitem[Yusef-Zadeh et al.(1996)]{Yusef96} Yusef-Zadeh, F., Roberts, D.A., Goss, W.M., Frail, D.A., Green, A.J. 1996, ApJ, 466, L25
\bibitem[Zinchenko et al.(2000)]{Zinchen00} Zinchenko, I., Henkel, C., \& Mao, R.Q. 2000, \aap, 361, 1079
%%\bibitem[Zylka et al.(1990)]{Zylka90} Zylka, R., Mezger, P.G., Wink, J.E. 1990, A\&A, 234, 133
\end{thebibliography}
\end{document}